\documentclass[twocolumn,showpacs,preprintnumbers,amsmath,amssymb,unsortedaddress]{revtex4}

\usepackage{graphicx}
\usepackage{dcolumn}
\usepackage{bm}

\def\bk{{\bf k}}

\def\bx{{\bf x}}

\def\by{{\bf y}}

\def\bq{{\bf q}}

\def\g{{\rm g}}

\def\abc2{{\left(\frac{ab}{c^2}\right)}}

\def\nbarg{\bar{n}_{\rm g}}

\def\dr{{d^3\!r}}

\def\dk{{d^3\!k}}
\def\dy{{d^3\!y}}

\def\be{\begin{equation}}
\def\ee{\end{equation}}
\def\rhob{\bar{\rho}}

\def\nbar{\bar{n}}
\def\simless{\mathbin{\lower 3pt\hbox
  {$\rlap{\raise 5pt\hbox{$\char'074$}}\mathchar"7218$}}}   
\def\simgreat{\mathbin{\lower 3pt\hbox
   {$\rlap{\raise 5pt\hbox{$\char'076$}}\mathchar"7218$}}}  

\def \Lin{\rm Lin}
\def \NL{\rm NL}
\def\ba{\begin{eqnarray}}
\def\ea{\end{eqnarray}}

\def\hc{\rm h}
\def\vir{\rm vir}
\def\bess0{{\rm J_0}}
\def\sbess0{{\rm j_0}}
\def\cyc{{\rm cyc}}
\def\1H{1{\rm H}}
\def\2H{2{\rm H}}
\def\3H{3{\rm H}}

\def\Mpc{{\rm Mpc}}
\def\Mpch{{h^{-1}\rm Mpc}}

\def\Fsym{F^{(\rm s)}}
\def\Gsym{G^{(\rm s)}}

\def\D1{\rm 1D}

\def\Loop{\rm 1Loop}
\def\cent{\rm c}

\def\MNRAS{{Mon. Not. R. Astron. Soc.~}}
\def\PRD{{Phys. Rev. D.~}}
\def\PRL{{Phys. Rev. Lett.~}}
\def\ApJ{{Astrophys. J.~}}
\def\AA{{Astron. Astrophys.~}}
\def\Nat{{Nature (London)~}}
\def\AstroPart{{Astro-particle Phys.~}}

\begin{document}

\title{Scale dependence of halo and galaxy bias: effects in real space}

\author{Robert E. Smith}
\affiliation{Department of Physics and Astronomy,
University of Pennsylvania, 209 South 33rd Street, 
Philadelphia, PA 19104, USA.}
\email{res@astro.upenn.edu}

\author{Rom\'an Scoccimarro} \affiliation{Center for Cosmology and
Particle Physics, Department of Physics, New York University, New
York, NY 10003, USA.}  \email{rs123@nyu.edu}

\author{Ravi K. Sheth}
\affiliation{Department of Physics and Astronomy, 
University of Pennsylvania, 209 South 33rd Street, 
Philadelphia, PA 19104, USA.}
\email{shethrk@physics.upenn.edu}

\date{\today}


\begin{abstract}
We examine the scale dependence of dark matter halo and galaxy
clustering on very large scales ($0.01 <k[h\,\Mpc^{-1}] <0.15$), due
to non-linear effects from dynamics and halo bias.  We pursue a two
line offensive: high resolution numerical simulations are used to
establish some old and some new results, and an analytic model is
developed to understand their origins. Our simulations show: (i) that
the $z=0$ dark matter power spectrum is suppressed relative to linear
theory by $\sim5\%$ on scales $0.05<k[h\Mpc^{-1}]<0.075$; (ii) that,
indeed, halo bias is {\em non-linear} over the scales we probe and
that the scale dependence is {\em a strong function of halo mass}.
High mass haloes show no suppression of power on scales
$k<0.07[h\Mpc^{-1}]$, and only show amplification on smaller scales,
whereas low mass haloes show strong, $\sim5-10\%$, suppression over
the range $0.05 <k[h\,\Mpc^{-1}] <0.15$.  These results were primarily
established through the use of the cross-power spectrum of dark matter
and haloes, which circumvents the thorny issue of shot-noise
correction. The halo-halo power spectrum, however, is highly sensitive
to the shot-noise correction; we show that halo exclusion effects make
this {\em sub-Poissonian} and a new correction is presented.  Our
results have special relevance for studies of the baryon acoustic
oscillation features in the halo power spectra.  Non-linear mode-mode
coupling: (i) damps these features on progressively larger scales as
halo mass increases; (ii) produces small shifts in the positions of
the peaks and troughs which depend on halo mass.  We show that these
effects on halo clustering are important over the redshift range
relevant to such studies $(0<z<2)$, and so will need to be accounted
for when extracting information from precision measurements of galaxy
clustering.

Our analytic model is described in the language of the `halo-model'.
The halo-halo clustering term is propagated into the non-linear regime
using `1-loop' perturbation theory and a non-linear halo bias model.
Galaxies are then inserted into haloes through the Halo Occupation
Distribution. We show that, with non-linear bias parameters derived
from simulations, this model produces predictions that are
qualitatively in agreement with our numerical results.  We then use it
to show that the power spectra of red and blue galaxies depend
differently on scale, thus underscoring the fact that proper modeling
of nonlinear bias parameters will be crucial to derive reliable
cosmological constraints. In addition to showing that the bias on very
large scales is not simply linear, the model also shows that the
halo-halo and halo-dark matter spectra do not measure precisely the
same thing.  This complicates interpretation of clustering in terms of
the stochasticity of bias.  However, because the shot noise correction
is non-trivial, evidence for this in the simulations is marginal.
\end{abstract}


\maketitle


\section{Introduction}

Statistical analysis of the large scale structures observed in galaxy
surveys can provide a wealth of information about the cosmological
parameters, the underlying mass distribution and the initial
conditions of the Universe. The information is commonly extracted
through measurement of the two-point correlation function
\cite{Hawkinsetal2003,Eisensteinetal2005} or its Fourier space
analogue the Power spectrum
\cite{Tegmarketal2004,Coleetal2005,Tegmarketal2006,Percivaletal2006a,Percivaletal2006b}.
When further combined with high precision measurements of the
temperature anisotropy spectrum from the Cosmic Microwave Background
very strong constraints can be imposed on the initial conditions, the
energy content, shape and evolution of the Universe
\cite{Spergeletal2006}.

For homogeneous and isotropic Gaussian Random fields, such as is
supposed for the post inflationary density field of Cold Dark Matter
(hereafter CDM) fluctuations, each Fourier mode is independent, and
thus all of the statistical properties of the field are governed by
the power spectrum. However, non-linear evolution of matter couples
the Fourier modes together, and power is transfered from large to
small scales \cite{Peebles1980,Bernardeauetal2002}.  Consequently, it
is non-trivial to relate the observed structures to the physics of the
initial conditions. Further, since one typically measures not the
mass, but the galaxy fluctuations, some understanding of the mapping
from one to the other is required.  This mapping, commonly referred to
as galaxy bias, encodes the salient physics of galaxy formation.

One last complication must be added: since galaxy positions are
inferred from recession velocities using Hubble's law, and because
each galaxy possesses its own peculiar velocity relative to the
expansion velocity, a non-trivial distortion is introduced to the
clustering on all scales from the velocity field. These velocity
effects are commonly referred to as redshift space distortions. Thus
one must accurately account for non-linear evolution of matter
fluctuations, bias and redshift space distortions in order to extract
precise information from large scale structure surveys and this
remains one of the grand challenges for modern physical cosmology.

In this paper, we investigate the issue of bias in some detail,
through both numerical and analytic means.  We focus on real space
effects and reserve our results from redshift space for a subsequent
paper. Our numerical work focuses on the generation of multiple
realizations of the same cosmological model, in two different box
sizes. This allows us to construct halo catalogues spanning a large
dynamic range in mass that are largely free from discreteness
fluctuations and the multiple realizations allow us to derive errors
that are `true errors from the ensemble'.  We use this data to show
that not only is halo clustering on very large scales scale dependent,
but that the scale dependence is a strong function of halo mass. These 
results are completely expected given the standard theoretical understanding 
of dark matter haloes based on the `peak-background split' 
argument~\cite{ColeKaiser1989,MoWhite1996,MoJingWhite1997,ShethTormen1999,
Scoccimarroetal2001}.

Our analytic approach to modelling these trends can be summarized as
follows.
\begin{itemize}
\item 
Haloes are biased tracers of the mass distribution.  To describe this
bias, we assume that the bias relation between the halo density field
and the dark matter field is non-linear, local and deterministic.
This allows us to use the formalism of
\cite{FryGaztanaga1993,MoJingWhite1997}.  In order for the local model
to hold, one must integrate out small scales where locality is almost
certainly violated. This can be done in real space by smoothing
small-scale fluctutations, or in Fourier space by considering small
wavenumbers 
\footnote{Several broad classes of bias model may be defined: local
\cite{Coles1993} and non-local
\cite{Kaiser1984,DekelRees1987,Catelanetal1998,Catelanetal2000,Matsubara1999};
linear or non-linear
\cite{CenOstriker1992,FryGaztanaga1993,Heavensetal1998,Mannetal1998,Bensonetal2000,PeacockSmith2000,CenOstriker2000};
and deterministic or stochastic
\cite{ScherrerWeinberg1998,DekelLahav1999}. With the exception of
stochastic, local, linear biasing, all these prescriptions result in
some non-trivial degree of scale dependence. Note that when weakly
nonlinear scales are discussed, as we do in this paper, it may not be
entirely consistent to neglect nonlocality, which could potentially
alter the predictions we present. Nonlocal bias can be looked for in
observations by using higher-order statistics
\cite{FriemanGaztanaga1994,Feldmanetal2001}. However, we may take
comfort in the fact that there is always a strong correlation between
haloes and dark matter, since the former are built from the latter!}.
\item The underlying CDM density field is then propagated into the
non-linear regime using standard Eulerian perturbation theory
techniques \cite{Bernardeauetal2002}.
\item Galaxies are then assumed to form only in haloes above a given
mass \cite{WhiteRees1978} and these are inserted into each halo using
the `halo model' approach
\cite{Bensonetal2000,PeacockSmith2000,Seljak2000,Scoccimarroetal2001,BerlindWeinberg2002}.
\end{itemize}
Because we write the PT evolved halo density field as a series
expansion we refer to this method as `Halo-PT' theory.

Our results are particularly relevant for studies which intend to use
the baryon acoustic oscillation feature (hereafter BAO) in the low
redshift clustering of galaxies to derive constraints on the dark
energy equation of state \cite{Eisensteinetal2005}.  The CDM transfer
function that we have adopted throughout contains a significant amount
of BAOs, and we give an accounting of the possible non-linear
corrections from mass evolution and biasing that might influence the
detection and interpretation of such features.  Previous work in this
direction has primarily focused on analysis of numerical simulations
\cite{Meiksinetal1999,SeoEisenstein2003,SeoEisenstein2005,White2005,Huffetal2006},
although several analytic works have recently been presented:
\cite{CrocceScoccimarro2006c} derive the exact damping of BAOs in the
Zel'dovich approximation and calculate it in the exact dynamics by
re-summing perturbation theory; \cite{JeongKomatsu2006} consider real
space corrections to the power spectrum from one-loop PT;
\cite{GuzikBernsteinSmith2006} use the halo model, also in real space,
to explore systematics; \cite{Eisensteinetal2006a,Eisensteinetal2006b}
consider a model of Lagrangian displacements fit to simulations.

In Section~\ref{sec:BAOs} we discuss how the approach we have
developed here is complementary to and expands on these studies.  In
Section~\ref{sec:N-Body} we discuss the numerical simulations and
present our measurements of scale dependence in the dark matter, halo
center and halo-dark matter cross power spectra.  We also present the
evidence for large scale non-linear bias.  Then in Section
\ref{sec:theory}, we outline some key notions concerning the halo
model of large scale structure, as this is the frame work within which
we work. In Section \ref{sec:halobias} we describe the non-linear bias
model that we employ. In Section \ref{sec:HaloPT} we use the 3rd Order
Eulerian perturbation theory to describe the evolved Eulerian density
field in terms of the inital Lagrangian fluctuations.  In Section
\ref{sec:HaloPow}, we use the 3rd order halo density fields to produce
an anlytic model for the `1-Loop' halo and halo-cross dark matter
power spectra. In Section \ref{sec:results} we explore the predictions
of the analytic model for a range of different halo masses. In Section
\ref{sec:comparison} we compare our analytic model to the nonlinear
bias seen in the numerical simulations.  We use the analytic model to
examine the galaxy power spectrum in Section~\ref{sec:galaxies}, and
present our conclusions in Section~\ref{sec:conclusions}.

Throughout, we assume a flat Friedmann-Lema\^itre-Robertson-Walker
(FLRW) cosmological model with energy density at late times dominated
by a cosmological constant ($\Lambda$) and a sea of collisionless cold
dark matter particles as the dominant mass density.  We take
$\Omega_m=0.27$ and $\Omega_{\Lambda}=0.73$, where these are the
ratios of the energy density in matter and a cosmological constant to
the critical density, respectively. We use a linear theory power
spectrum generated from {\ttfamily cmbfast}
\cite{SeljakZaldarriaga1996}, with baryon content of $\Omega_b =0.046$
and $h=0.72$. The normalization of fluctuations is set through
$\sigma_8=0.9$, which is the initial value of the r.m.s. variance of
fluctuations in spheres of comoving radius $8 h^{-1}\,\Mpc$
extrapolated to $z=0$ using linear theory.


\section{Motivation}\label{sec:BAOs}

A number of recent papers have attempted to quantify the scale
dependence of galaxy bias.  A subset of these have forwarded simple
analytic models to remove the scale-dependent biases in the power
spectrum estimator.  We discuss some of these below so as to set the
stage for our work.

%
\subsection{Cole et al. (2005)}
Based on analysis of mock galaxy catalogues from the Hubble volume
simulation, these authors proposed a simple analytic model to account
for the non-linear scale dependence:
\be P_{\g\g}(k)=b^2 P_{\Lin}(k)\frac{1+A_2k^2}{1+A_1k} 
\label{eq:cole} .\ee
The parameter $A_1=1.4$, and $A_2$ was allowed to vary over a narrow
range, which was then marginalized-over in the fitting procedure.
When $A_1k\ll1$, this model has
\be P_{\g\g}(k)\approx b^2 P_{\Lin}(k)\left[1-A_1k+A_2k^2\right] \
. \ee
We show below that the bracketed terms are suggestive of the
$P^{\delta\delta}_{13}+P^{\delta\delta}_{22}$ terms from the dark
matter perturbation theory, but with incorrect dependence on $k$. In addition, ignoring the fact that $A_1$ may
depend on galaxy type is a serious inconsistency.  For instance, our
results indicate that $A_1$ for LRG-like galaxies is smaller than 1.4.

A further concern regarding this model is that no accounting for
non-Poisson shot noise has been made. If galaxy formation takes place
only in haloes, then galaxies are not Poison samples of the mass
distribution. The analytic model we develop shows that it is important
to account for this, and how.  We are therefore skeptical about the
blind use of equation (\ref{eq:cole}), particularly with regard to its
use in the analysis of LRGs
\cite{Padmanabhanetal2006,Tegmarketal2006}.


\subsection{Seo \& Eisenstein~(2005)}

These authors examined the scale dependence of halo bias in a large
ensemble of low-resolution numerical simulations.  They proposed
\be P_{\g\g}(k)=b^2P_{\Lin}(k)+A_1k+A_2k^2+A_0,\ee
which bares some similarity to that of Cole et al., as can be seen by
re-writing $A_1\to A_1/(b^2P_{\Lin})$ and $A_2\to A_2/(b^2P_{\Lin})$.
The effective spectral index of the linear power spectrum evolves from
$0>n_{\rm eff}>-1$ over the range of $k$ of interest, so we can think
of $P_{\Lin}$ as being approximately constant.  Then, the inclusion of
the $b^2$ term decreases the effective $A_1$ as required, but it also
decreases the effective $A_2$; our results indicate that this is
inappropriate.

There is an important difference between this model and the previous
one---the inclusion of the constant power term $A_0$.  This was
introduced to account for `anomalous power' -- by which was meant
effects envisaged by \cite{ScherrerWeinberg1998} -- and or non-Poisson
shot noise following \cite{Seljak2001}.  Our analyis strongly supports
the inclusion of this term.


\subsection{Seljak~(2001); Schulz \& White~(2006);\\ Guzik, Bernstein \& Smith~(2006)}

These authors explored the scale dependence of galaxy bias in the halo
model focusing their attention on the 1-Halo term. They showed that if
this was taken simply as a non-Poisson shot-noise correction, then it
would be a significant source of scale dependence in the large scale
galaxy power spectrum. We have confirmed this in our study. These
studies lead one to suggest a base form of the kind:
\be P_{\g\g}(k)=b^2P_{\Lin}(k)+A_0\ .\ee
On top of this base form we need to include modifications that
capture the true non-linear evolution of the halo field.


\subsection{Huff et al.~(2006)}
These authors used a set of three large cosmological simulations to
investigate the scale dependence of the BAOs. They suggested
\be P_{\g\g}(k)= b^2P_{\Lin}(k) \exp\left[-\left(A_1k\right)^2\right]
    + A_0, \ee
where the exponential damping term was introduced to account for halo
profiles. Again examining the large scale behaviour, $A_1k\ll1$, we
see that this equation may be re-written
\be P_{\g\g}(k)\approx b^2P_{\Lin}\left[1-(A_1k)^2\right]+A_0.\ee
If $0<A_1< 1$ (the range they considered), then this formula will
suppress the power spectrum on scales $k<0.1 h\Mpc^{-1}$ by a percent
at most.  If this model is to account for the non-linear corrections
that we see, then $A_1>1$.  However, the strong exponential damping
makes it unlikely that this model will properly characterize the
transition from the 2- to 1-Halo term.  This is because, as we show
below, the nonlinear evolution of halo centers includes an additional
boost at intermediate $k$ which this model does not capture.


\subsection{A necessary model}

Our results suggest that a necessary model will have the following
properties: the model should be able to produce a pre-virialization
feature and a small scale nonlinear boost with $k$-dependencies
motivated by physical arguments; nonlinear corrections should depend
on galaxy type; a constant power term should be added to account for
non-Poisson shot noise. We therefore expect a reasonable starting
point for any empirical modelling of the large scale, scale dependence
of the galaxy power spectrum to be,
\ba 
\!\!P_{\g\g}(k,T) & = & b^2(T)\left[P_{\Lin}(k){\rm e}^{- A_1(T)k^2}
+A_2(T)\, k^{m(T)}\right] \nonumber \\ & & \times |W(k/k_{\alpha})|
+ A_0(T) |W(k/k_{\beta})|+\frac{1}{\nbar_{\g}}
\ .\ea
Our notation makes explicit that the coefficients have the following
properties: $b>0$, $A_0>0$, $A_1>0$, $A_2>0$ and $m\ge0$, and that all
depend on galaxy type $T$. The first term is composed of two pieces:
in the first piece, we have modeled the damping of BAOs using a
Gaussian as derived
in~\cite{CrocceScoccimarro2006a,CrocceScoccimarro2006b} for the dark
matter case; for the second piece, we have added a $k$-dependent boost
that models the power added by mode-mode coupling and nonlinear
bias. Our simple power-law form (with two parameters $A_2,m$) is meant
to describe this effect over a restricted range of scales. The fact
that this term is additive as opposed to multiplicative, is meant to
emulate the fact that in PT this corresponds to the
$P^{\delta\delta}_{22}$ term which arises from the convolution of
linear power on different scales and is therefore smooth possesing no
information on BAOs.  For weakly nonlinear scales it has a positive
spectral index (note, however, that in the limit $k\rightarrow0$,
$m=4$ is expected from momentum conservation arguments).  The second
term corresponds to Poisson shot noise from unequal weighting of
haloes. The last term corresponds to the Poisson shot noise from the
galaxy point distribution. We have included filter terms
$W(k/k_{\alpha})$ and $W(k/k_{\beta})$ to indicate the damping due to
density profiles, which will occur for $k>1/{r_{\rm vir}}$. This
function may be greatly simplified by examinig the case $k\ll 1
[h\Mpc^{-1}]$, for which it reduces to
\ba 
P_{\g\g}(k,T) & = & b^2(T)\left\{P_{\Lin}\left[1- A_1(T)k^2\right]
+A_2(T)\, k^{m(T)}\right\} \nonumber \\ & & 
+ \tilde{A}_{0}(T) , \hspace{0.5cm} \left(k\ll1\right)
\ ,\ea
where the parameter $\tilde{A}_0(T)$ subsumes all sources of constant
large scale power.


\section{Scale dependent halo bias from Numerical simulations}
\label{sec:N-Body}

\subsection{Simulation details and halo catalogues}

We have performed a series of high-resolution, collisionless dark
matter $N$-body simulations, where $N=512^3$ equal mass particles.
Each simulation was performed using {\tt Gadget2} \cite{Springel2005};
the internal parameter settings can be found in Table~1 of
\cite{CroccePeublasScoccimarro2006}, where more details about the runs
themselves are available.  The initial conditions were set-up using
the 2nd-order Lagrangian Perturbation theory at redshift $z=49$
\cite{CroccePeublasScoccimarro2006}, with linear theory power spectrum
taken from {\tt cmbfast} \cite{SeljakZaldarriaga1996}, with the
cosmological model being the same as that used throughout this paper.
We will present results from two different box sizes: 20 smaller
higher resolution box (hereafter HR) for which the volume is $V = L^3
= (512 \Mpch)^3$, and 8 realizations of a larger, lower resolution box
(hereafter LR) for which $L=1024 \Mpch$ box.

Haloes were identified in the $z=0$ outputs using the
friends-of-friends algorithm with linking-length parameter $l=0.2$.
Halo masses were corrected for the error introduced by discretization
of the halo density structure \cite{Warrenetal2005}.  Since the error
in the estimate of the halo mass diverges as the number of particles
sampling the density field decreases, we only study haloes containing
50 particles or more.  For the HR and LR simulations this corresponds
to haloes with $M>4.0 \times10^{12} h^{-1}M_{\odot}$ and $M>4.0
\times10^{13} h^{-1}M_{\odot}$, respectively.  We then constructed
four non-overlapping sub-samples of haloes with roughly equal numbers
per sub-sample.  The two low mass bins were harvested from the HR
simulations and the two high mass ones were taken from the LR
runs. Further details may be found in Table~\ref{table:halocat}.


\begin{table}
\caption{\small Halo samples. $N_{\rm real}$ is the number of
independent realizations. $\bar{N}_{\rm h}$ and $\nbar_{\rm h}$ are
the ensemble average number and number densities of haloes in each
mass bin.}\label{table:halocat}
\begin{ruledtabular}
  \begin{tabular}{llcccc}
  & & $N_{\rm real\!\!}$ & $L \, [\Mpc\, h^{-1}]$ & $\bar{N}_{\rm h}\ [\times10^4]$
          & $\nbar_{\rm h}\ [\Mpc^{-3} h^{3}]$\\ \hline
LR & Bin 1 
\footnote{Mass bin 1 = $ M>1.0\times10^{14}h^{-1}M_{\odot}$} 
         & 8 & 1024 & $3.6863$ & $3.43313\times10^{-5}$\\
LR & Bin 2 
\footnote{
Mass bin 2 = $1.0\times10^{14}h^{-1}M_{\odot}>
M>4.0\times10^{13}h^{-1}M_{\odot}$} 
         & 8 & 1024 & $7.3530$ & $6.8480\times10^{-5}$\\
HR & Bin 3 
\footnote{Mass bin 3 = $ 4.0\times10^{13}h^{-1}M_{\odot}>
M>7.0\times10^{12}h^{-1}M_{\odot}$}
         & 20 & 512 & $6.9287$ & $5.1623\times10^{-4}$\\
HR & Bin 4 
\footnote{
Mass bin 4 = $ 7.0\times10^{12}h^{-1}M_{\odot}>
M>4.0\times10^{12}h^{-1}M_{\odot}$}
         & 20 & 512 & $5.5415$ & $4.1287\times10^{-4}$\\
\hline 
\end{tabular}
\end{ruledtabular}
\end{table}


\begin{figure*}
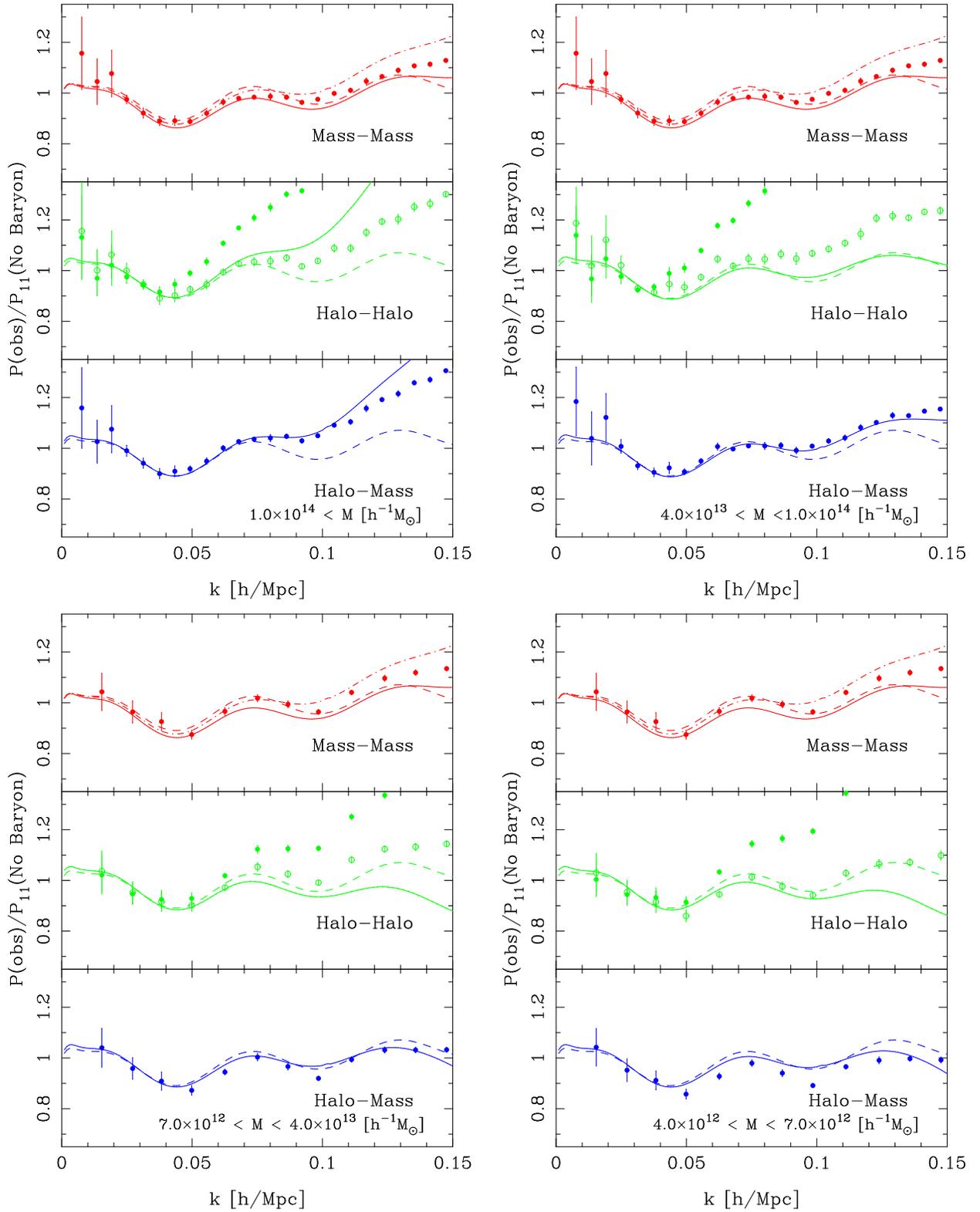

\centerline{
\includegraphics[width=8cm]{FIG.1a.ps}
\hspace{0.3cm}
\includegraphics[width=8cm]{FIG.1b.ps}}
\vspace{0.2cm}
\centerline{
\includegraphics[width=8cm]{FIG.1c.ps}
\hspace{0.3cm} 
\includegraphics[width=8cm]{FIG.1d.ps}}
\caption{\small{Halo power spectrum measurements and predictions,
ratioed with a smooth `No Baryon' dark matter power spectrum, for four
bins in halo mass: results for massive haloes (the top two panels) are
from the LR simulations, whereas lower masses (the two bottom panels)
are from the HR simulations.  Filled circles in the top, middle and
bottom sections of each panel show the ensemble average non-linear
$P^{\delta\delta}(k)$ $P^{\hc\hc}(k)$, and $P^{\delta\hc}$,
respectively.  The open circles in the middle sections show
$P^{\hc\hc}(k)$ with the non-standard shot noise subtraction described
in Appendix~ \ref{app:HaloDiscrete}.  In all panels the linear theory
dark matter, halo-halo and halo-cross power spectra are shown as
dashed lines.  The top panel also shows predictions from {\tt halofit}
(solid lines) \cite{Smithetal2003} and 1-Loop perturbation theory
(dot-dash lines). Solid lines in the middle and bottom panels show our
new analytic model, Halo-PT.
\label{fig:romansimHaloPT} }}
\end{figure*}


\subsection{Mass, halo and halo-Mass power spectra}
\label{ssec:halopowerspectra}

For each realization and each bin in halo mass we measured the
following quantities: the power spectrum of the dark matter
$P^{\delta\delta}(k)$; the power spectrum of dark matter haloes
$P^{\hc\hc}(k)$; and the cross power spectrum of dark matter and dark
matter haloes, $P^{\delta\hc}(k)$.  The power spectra may be generally
defined
\be \left<\delta^{\alpha}(\bk)\delta^{\beta}(\bk')\right>\equiv
(2\pi)^3\delta^{D}(\bk+\bk') P^{\alpha\beta}(k) \ ,\ee
where
\be 
\delta^i(\bk)\equiv \left[
\begin{array}{c}
\delta(\bk)\\
\delta^{\hc}(\bk)
\end{array}\right]
\ ;\ \ 
P^{ij}(k)\equiv\left[
\begin{array}{cc}
P^{\delta\delta}(k) & P^{\delta\hc}(k)\\
P^{\hc\delta}(k) & P^{\hc\hc}(k)
\end{array}\right]
\ ,\ee
with $\delta(\bk)$ and $\delta^{\hc}(\bk)$ being the Fourier
transforms of the mass and halo density perturbation fields, 
\be \rho^{i}(\bx)=\rhob^{i}\left[1+\delta^{i}(\bx)\right],\ee
where the index $i$ again distinguishes between dark matter and
haloes. e.g. $\rhob^1\equiv \rhob$ and $\rhob^2\equiv\rho^{\hc}$.

We estimate the spectra through the conventional Fast Fourier
Transform (FFT) method \footnote{The dark matter particles/halo centers were assigned to a
regular cubical grid using a fourth-order interpolation scheme, and
each point on the grid was given equal weight. The FFT of the gridded
density field was then computed. Each resulting Fourier mode was
corrected for convolution with the grid by dividing by the Fourier
transform of the mass assignment window function.  The power spectra
on scale $k_l$ are then estimated by performing the following sums,
\[\hat{P}^{ij}(k_l)=\frac{1}{M}\sum_{l=1}^{M}\left|\delta^{i}(\bk_l)
(\delta^j(\bk_l))^*\right|^2 \, ,\]
where $M$ are the number of Fourier modes in a spherical shell in
$k$-space of thickness $\Delta k$, and $*$ denotes complex
conjugation.}  (for a detailed discussion see \cite{Smithetal2003}).
The mean power and 1-$\sigma$ errors on the spectra were estimated
from the ensemble neglecting the bin-to-bin covariances.  Inspection
of an estimate of the covariance matrix from the 20 HR simulations
showed that this is reasonable. There is, however, a small degree of
off diagonal covariance, but the number of simulations was
insufficient to make a precise estimate. 

Figure~\ref{fig:romansimHaloPT} shows the three types of power spectra
measured from the ensemble of simulations for each of the four bins in
halo mass described in Table~\ref{table:halocat}.  The filled circles
and associated error bars in the top, middle and bottom sections of
each panel show $\hat{P}^{\delta\delta}(k)$, $\hat{P}^{\hc\hc}(k|M)$,
and ${\hat P}^{\delta\hc}(k|M)$.  The open circles show the result of
applying a non-standard shot-noise correction to $\hat{P}^{\hc\hc}$,
which we describe in Appendix~\ref{app:HaloDiscrete}.

To emphasize the non-linear evolution of the spectra and the BAOs in
particular, we have divided each spectrum by a smooth linear theory
spectrum, which we shall refer to as our `No Baryon' model.  This was
constructed by performing a chi-squared fit of the {\tt cmbfast}
transfer function data to the smooth transfer function model of
\cite{BondEfstathiou1984},
\be T(k)=\left[1+\left\{aq+(bq)^{1.5}+(cq)^2\right\}^d\right]^{-1/d}.
\label{eq:NoBaryon}\ee
The derived parameters are: $q=k/0.19$, $a=4.86$, $b=4.81$, $c=1.72$,
$d=1.18$.

To compare the halo spectra with the linear theory we require
estimates of the halo bias on very large scales, which we measure as
follows.  We begin by assuming that the spectra can be written
\be P^{ij}(k|M)={\mathcal A^{ij}}P_{\Lin}(k)\ ;\ {\mathcal
A}^{ij}=\left[
\begin{array}{cc} 
1 & b^{\hc\delta} \\
b^{\delta\hc} & (b^{\rm hh})^2,
\end{array}
\right]\label{eq:amplitude}\ee
where $i$ and $j$ denote the type of spectrum considered and where,
for reasons that will later be apparent, we distinguish between the
bias from $P^{\delta\hc}$ and $P^{\hc\hc}$. Hence, the likelihood of
obtaining an estimate of the power $\hat{P}^{ij}$ in the $l^{\rm th}$
$k$-bin is assumed to be an independent Gaussian with dispersion
$\sigma_i$
\be {\mathcal L}(\hat{P}^{ij}_l,\sigma_l|{\mathcal A},P_{\Lin})
=\frac{1}{\sqrt{2\pi}\sigma_l}
\exp\left\{-\frac{[\hat{P}^{ij}_l-{\mathcal A}P_{\Lin}(k_l)]^2}
{2\sigma_l^2}\right\}\ .\ee
Thus the combined likelihood of obtaining the data set
$\left\{\hat{P}^{ij}_l,\sigma_l\right\}$ can be written,
\be {\mathcal L}(\{\hat{P}^{ij}_l\},\{\sigma_l\}|{\mathcal A},
            P_{\rm Lin}) =\prod_{l=1}^{N_{\rm dat}} {\mathcal L}
(\hat{P}^{ij}_i,\sigma_l|{\mathcal A},P_{\Lin})\ .\ \ee
On maximizing the likelihood function, we find the following estimator
for the halo bias matrix
\be \hat{\mathcal A}=\frac{\sum_{l=1}^{N_{\rm dat}}P_{\Lin}(k_l)
\hat{P}^{ij}_l/\sigma^2_l} {\sum_{l=1}^{N_{\rm
dat}}[P_{\Lin}(k_l)/\sigma_l]^2}\ .\label{eq:Aestimate}\ee
We construct error estimates through further differentiation of the
Gaussian likelihood function:
\be \sigma_{\mathcal A} = \left(\frac{\partial^2\log{\mathcal
L}}{\partial {\mathcal A}^2}\right)^{-1/2} = \left(\sum_{i=1}^{N_{\rm
dat}} \left[\frac{P_{\Lin}(k_i)}{\sigma_i}\right]^2\right)^{-1/2} \
.\ee
Lastly, since we observe scale-dependent non-linear effects in the
matter power spectrum for $k>0.05 h\,\Mpc^{-1}$, we only use modes
with $k<0.04 h\,\Mpc^{-1}$ in the fitting of the amplitude matrix
${\mathcal A}$. Estimates of the large scale bias parameters
$b^{\hc\hc}$ and $b^{\delta\hc}$ are presented in
Table~\ref{table:halobias}. \footnote{The procedure was also performed
separately for the no baryon model spectrum, this was done in order
for ratios to be taken.}

The dark matter power spectra in Fig.~\ref{fig:romansimHaloPT} show
significant deviations away from linear theory prediction: at $0.05<k
[h\Mpc^{-1}]<0.075$, there is a suppression of power relative to
linear theory, whereas at $k>0.075 [h\Mpc^{-1}]$ there is an
amplification.  Perturbation theory studies \cite{Bernardeauetal2002}
refer to the suppression effect (caused by tidal terms) as
`pre-virialization'.  Recently, this has been understood in much more
detail as a result of the damping of linear features by
nonlinearities, leading to an exponentially decaying propagator that
measures the loss of memory of the density field to the initial
conditions \cite{CrocceScoccimarro2006a,CrocceScoccimarro2006b}.
Although the effect in the power is rather well-known and has been
observed in recent numerical simulations of CDM
spectra\cite{Percivaletal2001,Smithetal2003,
Coleetal2005,Springeletal2005}, our results constitute a rather
precise measurement of this effect, with realistic errors drawn from
the ensemble. A complete assessment of the damping of linear features
such as BAOs is done by studying the
propagator~\cite{CrocceScoccimarro2006a,CrocceScoccimarro2006b} rather
than the power spectrum. See~\cite{CrocceScoccimarro2006c} for further
discussion on this.

The solid and dot-dashed lines show predictions based on {\tt halofit}
\cite{Smithetal2003} and on 1-Loop perturbation theory (PT). 
The PT results do well compared to the simulations on very large
scales, but for $k>0.07$ they increasingly over-predict the power.  We
note also that PT appears to predict the pre-virialization feature in
the simulations, adding additional support to the claim that
$\hat{P}^{\delta\delta}(k)$ requires non-linear corrections on the
scales of interest. However, qualitatively, it under-predicts the
magnitude of the effect. We note that {\tt halofit} does reasonably
well at capturing the behavior for $k<0.07 h Mpc^{-1}$, but it appears
to under predict the measured data.

Before moving on to $P^{\hc\hc}$ and $P^{\delta\hc}$, we think it is
worth noting that the BAOs in $P^{\delta\delta}$ on $k>0.1 h\Mpc^{-1}$
have been erased.  The large-box LR measurements show that the third
peak is gone, and the height of the second peak has dropped so that it
appears more as a plateau. However, the behavior at $k<0.05
h\Mpc^{-1}$ appears unchanged.

Discreteness corrections for $P^{\delta\delta}(k)$ have been studied
in some depth\cite{Smithetal2003}.  However, for $P^{\hc\hc}$, the
appropriate correction is more complicated because haloes are rare,
highly clustered, and spatially exclusive.  In Appendix
\ref{app:HaloDiscrete} we show that the standard Poisson shot noise
correction for the cluster power spectrum results in {\em negative
power} at high $k$. This lead us to propose a new method for making
the `shot-noise' correction that accounts for exclusion, which we
discuss therein. The open circles in the middle sections of each panel
in Figure~\ref{fig:romansimHaloPT} show the result of this new
correction.  Filled circles show the uncorrected power, and stars show
the standard correction ---clearly, the choice of correction is
crucial. Note that owing to the arbitrary normalization things for the
standard shot noise method look better than they actually are.
Unfortunately, the residual uncertainties in our new procedure prevent
us from making strong statements about the scale dependence of
halo-halo clustering.

Whilst the discreteness correction is troublesome for $P^{\hc\hc}$ it
is almost negligible for $P^{\delta\hc}$ (the halo model arguments
which follow allow us to quantify this).  Our estimates of
$P^{\delta\hc}$ are shown in the bottom sections of the panels in
Fig~\ref{fig:romansimHaloPT}.  Notice that the scale dependence of
$P^{\delta\hc}$, is a strong function of halo mass.  $P^{\delta\hc}$
for the most massive haloes shows no deviations from linear theory
until $k>0.7 h\Mpc^{-1}$.  However, the pre-virialization feature
appears and gets enhanced as one goes to lower masses. Indeed, for our
lowest mass bin, $P^{\delta\hc}$ is sub-linear until $k>0.15
h\Mpc^{-1}$.

This has important consequences for the BAOs.  In the highest mass bin
(top left panel), the oscillations in $P^{\delta\hc}$ at
$k>0.07h\Mpc^{-1}$ have been erased.  However, the first trough, at
$k\sim0.04 h\Mpc^{-1}$, is unaffected.  For the next mass bin (top
right panel), the first peak and trough are unmodified, and that the
second peak is becoming noticeable.  This trend continues as we
decrease mass; there is even a hint of the third peak in the bottom
panels.  These
measurements indicate that non-linear dynamics can erase oscillations
on progressively larger scales as halo mass increases and that small
displacements to the positions of the peaks and troughs may occur and
that these will also be dependent on halo mass.  If the locations of
these peaks and troughs are to play an important role in constraining
cosmological parameters, our measurements suggest that understanding
and quantifying these displacements will be very important.

Before continuing, we comment on the possible explanation of
these shifts through simple scatter from cosmic variance. We remark
that it is certainly possible to reconcile some of the shifts in the
peak positions through this. However, we draw attention to the fact
that all of the points $k \ge 0.05$ in the cross-power spectra of the
low mass haloes are systematically lower than expected from the linear
theory.  We also re-iterate that the derived error bars are the errors
on the means for 20 realizations.  One caveat is that since the
spectra are normalized by the very-large scale modes of the power
spectrum, where cosmic variance errors are larger, we expect some
small fluctuations in the relative amplitudes of the theory
predictions as more data is acquired. Estimates of the error in the
present LR simulations suggest changes of the order $\sim1\%$ to be
acceptable; and this increases to $\sim3\%$ for the HR simulations.
If the amplitudes for the theory curves are too low by $\sim3\%$, then
some of these discrepancies may be alleviated. However, it is
unquestionable that non-linear effects are present on these scales and
we must therefore firmly accept that it is likely that these may cause
some shifting of the harmonic series.  Only a wider and expanded
numerical study will be able to address and answer these questions
more completely.

The solid lines in the middle and bottom sections of each panel show
predictions from the analytic model described in the following
sections.  In all cases this model provides a better description than
does linear theory.


\begin{figure*}
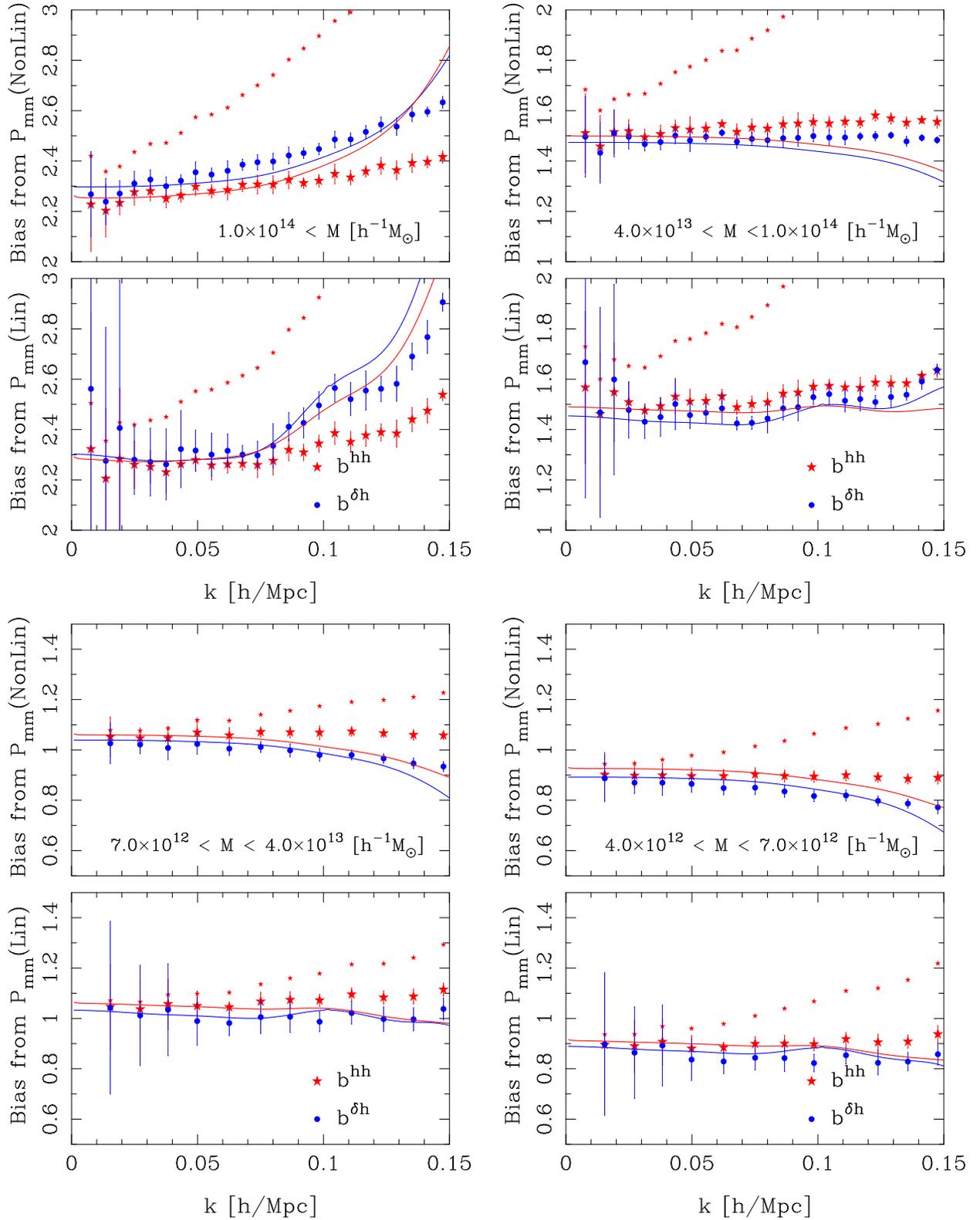

\centerline{
\includegraphics[width=8cm]{Fig.2a.ps}
\hspace{0.3cm}
\includegraphics[width=8cm]{Fig.2b.ps}}
\vspace{0.2cm}
\centerline{
\includegraphics[width=8cm]{Fig.2c.ps}
\hspace{0.3cm}
\includegraphics[width=8cm]{Fig.2d.ps}}
\caption{\small{Large scale halo bias derived directly from $N$-body
simulations for four bins in halo mass. The top sections of each panel
show the estimators $\hat{b}^{\delta\hc}_{\rm NL}$ (solid blue points)
and $\hat{b}^{\hc\hc}_{\rm NL}$ with and without shot noise correction
(large and small red stars, respectively). The bottom panels show the
same, but for $\hat{b}^{\delta\hc}_{\Lin}$ and
$\hat{b}^{\hc\hc}_{\Lin}$ (See equations \ref{eq:biasestimator1}
and \ref{eq:biasestimator2} for definitions). The solid blue and red
lines in each panel show the predictions for the bias from our Halo-PT
model.}
\label{fig:romansimbias}}
\end{figure*}


\begin{table}
\caption{\small Bias parameters for the halo samples.  $b^{\rm ST}_1$,
$b^{\rm ST}_2$ and $b^{\rm ST}_3$ are the first three non-linear halo
bias parameters derived from the Sheth \& Tormen model
\cite{ShethTormen1999,Scoccimarroetal2001} averaged over halo
bins. $b^{\delta}_1$, $b^{\delta}_2$ and $b^{\delta}_3$ are the
parameters measured from the $\delta^{\hc}$--$\delta$ scatter
plots. $b^{\delta\hc}$ is the large scale bias parameter measured
directly from $P^{\delta\hc}$.}\label{table:halobias}
\begin{ruledtabular}
  \begin{tabular}{lccccccc}
  & $b^{\delta\hc}$ & $b_1^{\rm ST} $ & $b_2^{\rm ST}$ &
$b_3^{\rm ST}$ & $b^{\delta}_1$ & $b^{\delta}_2$ & $b^{\delta}_3$ 
\\ 
\hline 
Bin 1 \footnote{See Table \ref{table:halocat} for definition of bins.}
              & $2.28\pm0.03$ & 2.19  & 0.94  & -0.98
              & 2.23 & 1.68 & -4.08 \\
Bin 2 
              & $1.49\pm0.04$ & 1.53  & -0.30  & 0.94 
              & 1.41 & 0.04 & -1.29 \\
Bin 3 
              & $1.02\pm0.03$ & 1.13  & -0.46  & 1.47 
              & 1.04 & -0.85 & 0.37 \\
Bin 4
              & $0.87\pm0.03$ & 0.98  & -0.44  &  1.44
              & 0.91 & -0.74 & 0.55 \\
\hline 
\end{tabular}
\end{ruledtabular}
\end{table}


\subsection{Scale dependence of the bias}\label{ssec:scaledebendence}

Next, we examine the scale dependence of halo bias.  We will consider
\be \hat{b}^{\hc\hc}_{\rm NL}(k_i)=
\sqrt{\frac{\hat{P}^{\hc\hc}(k_i)}{\hat{P}^{\delta\delta}(k_i)}} \ ;\
\ \hat{b}^{\delta\hc}_{\rm NL}(k_i)=
\frac{\hat{P}^{\delta\hc}(k_i)}{\hat{P}^{\delta\delta}(k_i)}\ .
\label{eq:biasestimator1}\ee
as well as 
\be \hat{b}^{\hc\hc}_{\Lin}(k_i)=
\sqrt{\frac{\hat{P}^{\hc\hc}(k_i)}{P_{\Lin}(k_i)}}\ ;\ \
\hat{b}^{\delta\hc}_{\Lin}(k_i)= \frac{\hat{P}^{\delta\hc}(k_i)}
{P_{\Lin}(k_i)}\ .
\label{eq:biasestimator2}\ee
For any particular realization the wave modes of the halo and dark
matter density fields are almost perfectly correlated.  Because the
first set of estimators are derived from taking the ratio of measured
power spectra, they are insensitive to this source of cosmic variance.
In this sense, the second set of estimators are non-optimal.  However,
they are the ones which will be used with real data, since
$P^{\delta\delta}$ is generally not observable.

In Figure \ref{fig:romansimbias} we show the results of measuring
these quantities for the same halo mass bins as in the previous
section.  The top and bottom parts of each panel show
(\ref{eq:biasestimator1}) and (\ref{eq:biasestimator2}), respectively.
The error bars, which were derived from the ensemble to ensemble
variations, are significantly larger for (\ref{eq:biasestimator2})
than for (\ref{eq:biasestimator1}), as expected.  The solid lines show
the predictions from the new analytic model described in the next
section.

For $b^{\hc\hc}$ we show both the shot-noise corrected (large stars)
and un-corrected (small stars) results. As was the case for the
halo-halo power spectra, we see that this correction is important, so
it must be known rather accurately.  The estimators for
$b^{\delta\hc}$ are also shown (filled circles).  Except for the
highest mass bin, $b^{\hc\hc}\ge b^{\delta\hc}$.  Indeed, as we shall
argue later, there are compelling theoretical reasons why the biases
derived from $P^{\delta\hc}$ and $P^{\hc\hc}$ are not in fact the
same, and that, one generally expects $b^{\hc\hc}\ge
b^{\delta\hc}$. However owing to the uncertainty regarding the shot
noise correction, no firm statement can be drawn from the current
data.

Two possible explanations why the highest mass halo bin appears to
behave differently are: Firstly, if the shot noise correction to
$P^{\hc\hc}$ is too aggressive, we may have underestimated the true
$P^{\hc\hc}$ and therefore the bias.  Alternatively, we have made no
shot-noise correction to $P^{\delta\hc}$; if one should be applied (we
think this is unlikely, owing to the large number of dark matter
particles), then our current estimate of $b^{\delta\hc}$ may be biased
high.

The estimators based on equation~(\ref{eq:biasestimator2}) show more
scale dependence than those based on~(\ref{eq:biasestimator1}),
especially for the two high mass bins.  As noted by
\cite{GuzikBernsteinSmith2006}, this is because the BAOs are erased on
larger and larger scales as higher and higher mass haloes are
considered.  Thus on dividing the halo spectra by a linear theory BAO
spectrum, we are in fact introducing scale dependence from the linear
model.

For the highest mass haloes, $\hat{b}^{\delta\hc}_{\rm NL}$ is
constant at $k<0.04h\Mpc^{-1}$, but it increases monotonically as $k$
increases.  This is a direct consequence of the absence of
pre-virialization in $P^{\hc\hc}$ on intermediate scales and the rapid
onset of non-linear power on smaller scales, compared to
$P^{\delta\delta}$.  The bias $\hat{b}^{\delta\hc}_{\rm NL}$ is flat
for haloes with masses
$(4.0\times10^{13}<M[h^{-1}M_{\odot}]<1.0\times10^{14})$, suggesting
that small clusters and group mass haloes are linearly biased tracers
of the non-linear dark matter.  The smallest two bins in halo mass
show the reverse trend: the bias decreases at large $k$, by $\sim10\%$
compared to the approximately constant value at smaller $k$.


\section{Theoretical Model}\label{sec:theory}

We now describe a model for interpreting the trends seen in the
previous section.  Our discussion is based on the `halo-model', which
we briefly summarize below.  See~\cite{CooraySheth2003} for a more
detailed review.


\subsection{The `Halo Model' of large scale structure}

The halo model may be described by the simple statement:
\begin{itemize}
\item All dark matter in the Universe is contained within a
distribution of CDM haloes, with masses drawn from some mass function
and with the density profile of each halo being drawn from some
universal stochastic profile.
\end{itemize}
The model attains its full potential when the second assumption is
stated:
\begin{itemize}
\item All galaxies exist only in isolated dark matter haloes, with
more massive haloes hosting multiple galaxies.
\end{itemize}
In essence the model has been in existence for several decades
\cite{NeymanScott1952,Peebles1974,McClellandSilk1977,WhiteRees1978,
ScherrerBertschinger1991,MoWhite1996,ShethJain1997}.  However it was
not until the advent of large numerical $N$-body simulations and
accurate characterization of halo phenomenology that its true value
was realized. Namely, given an appropriate Halo Occupation
Distribution (HOD -- i.e., a prescription for the number and spatial
distribution of galaxies within a halo) the model successfully
reproduces the real-space form of the two-point correlation function
of galaxies over a wide range of scales.  It predicts subtle
deviations from a power-law which have recently been seen in
observations \cite{Zehavietal2004} and provides a framework for
describing the luminosity \cite{Yangetal2004,Zehavietal2005} and
environmental dependence of galaxy clustering
\cite{Skibbaetal2006,AbbasSheth2006}. It also enables new tests of the
CDM paradigm to be constructed \cite{SmithWatts2005,SmithWattsSheth2006}.


\subsection{Power spectrum}

In the model the density fields of haloes and dark matter may be
written as a sums over haloes,
\be \rho^i(\bx)\equiv\rhob^i\left[1+\delta^i(\bx)\right]=
\sum_{j}^{N_i}M_j U_j(\bx-\bx_j|M_j)\ ,\ee
where $i=\{1,2\}$ distinguishes between haloes and dark matter and
$N_1=N$ and $N_2=N_{\hc}$ are the total number of haloes and the
number of haloes in some restricted range in mass. $M_j$ and $\bx_j$
are the mass and center of mass of the $j$th halo and
$U_j\equiv\rho_j(\bx_j)/M_j$ is the mass normalized density profile.
Following \cite{ScherrerBertschinger1991}, the power spectra
$P^{\delta\delta}$, $P^{\delta\hc}$ and $P^{\hc\hc}$ can be written as
the sum of two terms:
\be P^{ij}(\bk) = \frac{}{}P^{ij}_{\1H}(\bk)+P^{ij}_{\2H}(\bk)\ .
\label{eq:PowHaloModel}\ee
The first term, $P^{ij}_{\1H}$, referred to as the `1-Halo' term,
describes the intra-clustering of dark matter particles within single
haloes; the second, $P^{ij}_{\2H}$, referred to as the `2-Halo' term,
describes the clustering of particles in distinct haloes. They have
the explicit forms:
\ba P^{ij}_{\1H}(\bk) & = &
\frac{1}{\rhob^{i}\rhob^{j}}\int_{0}^{\infty} dM n(M) M^2 \left|
U(\bk|M)\right|^2 \nonumber\\ & & \times \Theta_{ij}(M,M)
\label{eq:Pow1H}\ ;\\
P^{ij}_{\2H}(\bk) & = & \frac{1}{\rhob^i\rhob^j}\int_{0}^{\infty}
\prod_{l=1}^{2} \left\{dM_l n(M_l) M_l U_l(\bk|M_l)\right\} \nonumber
\\ & & \times\ P^{\hc\hc}_{\cent}(\bk|M_1,M_2)\Theta_{ij}(M_1,M_2)\
,\label{eq:Pow2H} \ea
where $n(M)$ is the halo mass function, which gives the number density
of haloes with masses in the range $M$ to $M+dM$, per unit mass.  The
$\theta_{ij}$ matrix carves out the halo density field to be
considered, e.g. for haloes with mass $M>M_{\rm cut}$ the matrix is
\ba \Theta_{ij}(M_1,M_2) & \equiv & \nonumber \\ 
& & \hspace{-2.5cm}\left[
\begin{array}{cc}
\Theta(M_1)\Theta(M_2)  & \Theta(M_1)\Theta(M_2-M_{\rm cut}) \\
\Theta(M_1-M_{\rm cut})\Theta(M_2) & \Theta(M_1-M_{\rm cut})
\Theta(M_2-M_{\rm cut})
\end{array}
\right] \nonumber \ea
where $\Theta(x)$ is the Heaviside Step Function. More complicated
halo selections can easily be described through the $\Theta_{ij}$
notation. Lastly, $P^{\hc\hc}_{\cent}(\bk|M_1,M_2)$ is the power
spectrum of halo centers with masses $M_1$ and $M_2$.  This function
contains all of the information for the inter-clustering of haloes;
precise knowledge of this term is required to make accurate
predictions on large scales.

In principle, $P^{\hc\hc}_{\cent}(\bk|M_1,M_2)$ is a complicated
function of $M_1$, $M_2$ and $\bk$.  Initial formulations of the halo
model
\citep{Seljak2000,PeacockSmith2000,MaFry2000,Scoccimarroetal2001}
assumed that it could be well-approximated by
\be P^{\hc\hc}_{\cent}(k|M_1,M_2)=b_1(M_1)b_1(M_2)\,P_{\Lin}(k)\
,\label{eq:Lin2Halo}\ee
where all the scale dependence is in $P_{\Lin}(k)$, which is taken
from linear theory, and all the mass dependence is in the
scale-independent bias parameter $b_1(M)$
\cite{MoWhite1996,ShethTormen1999}.  In this approximation,
$P^{\hc\hc}_{\cent}(\bk|M_1,M_2)$ is a separable function of $M_1$,
$M_2$ and $\bk$.  As we show below, comparison with numerical
simulations shows that this simple model over-predicts power on very
large scales and provides insufficient power on intermediate scales.
In both cases, this is about a ten percent effect.

This discrepancy is not unexpected
\citep{Shethetal2001,BerlindWeinberg2002,YangMovandenBosch2003}.  A
simple correction results from setting
\be P^{\hc\hc}_{\cent}
(k|M_1,M_2)=b_1(M_1)b_1(M_2)P_{\NL}(k)\ ,\label{eq:NL2Halo}\ee
where $P_{\NL}$ is the non-linear rather than the linear matter power
spectrum, and, in addition, imposing an exclusion constraint:
\be \xi^{\hc\hc}_{\cent}(r|M_1,M_2)=-1 \ ; \ \
(r<r_{\vir_1}+r_{\vir_2})\ ,\ee
where $r_{\vir}$ is the virial radius of the halo. 

\begin{figure}
\centerline{
\includegraphics[width=8cm]{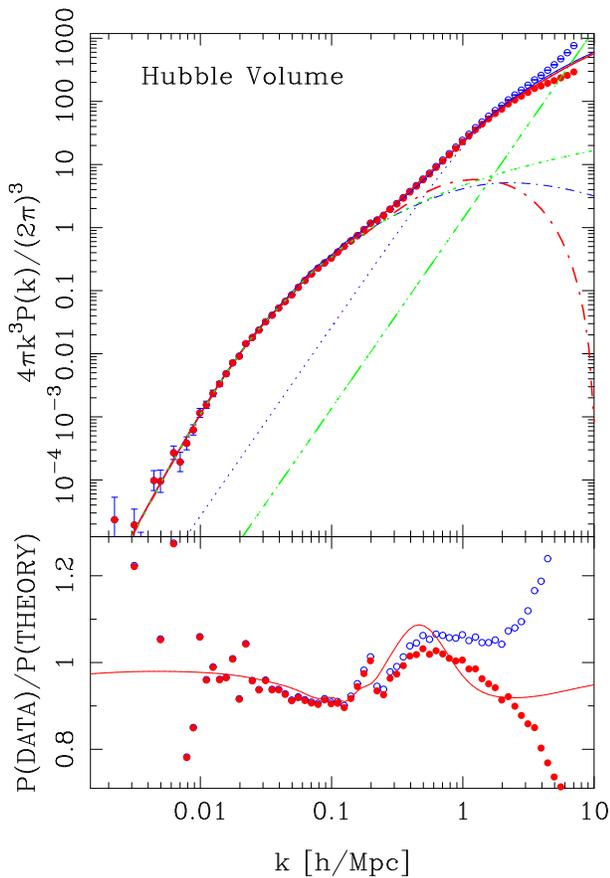}}
\caption{\small{Dark matter power spectrum measured as a function of
wavenumber measured from the $z=0$ time slice of the Hubble Volume
simulation \cite{Evrardetal2002}. In the top panel the red and blue
points show the estimates of the dark matter power spectrum measured
from the simulation, with and without a Poisson shot noise correction.
The dot-dash line shows the linear theory and the triple dot-dash line
shows the Poisson correction. The blue dotted and dot-dash curves show
the 1- and 2-Halo terms. The thick red dot-dash curve shows the 2-Halo
term where $P_{\rm NL}$ has been used instead of $P_{\Lin}$. The
bottom panel presents the ratio with respect to the halo model, but
with equation (\ref{eq:Lin2Halo}) for the 2-Halo term. The solid red
line shows the effect of the $P_{\rm NL}$ modification.
\label{fig:HubblePow}}}
\end{figure}

The success of this approach is demonstrated by comparing
$P^{\delta\delta}(k)$ measured in the $z=0$ output of the Hubble
Volume simulation \cite{Evrardetal2002} with the halo model
calculation.  The open and filled symbols in the top panel of
Figure~\ref{fig:HubblePow} show the measurement before and after
subtracting a Poisson shot noise term (which is shown by the triple
dot-dashed line.)  The dot-short dash line shows the linear theory
prediction, and the other two dot-dashed curves show two estimates of
the 2-halo term: the one which drops more sharply at large $k$ is
based on equation~(\ref{eq:NL2Halo}) and the other one is based on the
original approximation of equation~(\ref{eq:Lin2Halo}).  Since
equation~(\ref{eq:NL2Halo}) requires the use of a nonlinear power
spectrum, we used the one provided by \cite{Smithetal2003}.

The symbols in the bottom panel show the measurements divided by the
halo model calculation which uses equation~(\ref{eq:Lin2Halo}) for the
2-Halo term (i.e. the initial linear theory-based approximation).
Notice how they drop below unity at $k\sim 0.1h\Mpc^{-1}$.  The solid
line shows the halo model calculation which is based on
equation~(\ref{eq:NL2Halo})---it reproduces this pre-virialization
feature well.

As an interesting aside, we note that the fitting formula {\tt
halofit} does very well at matching the pre-virialization feature.
Whilst it is not apparent from the figure we also point out that the
transfer function of the Hubble volume simulation does contain BAOs;
thus, our results demonstrate that the fitting formula of
\cite{Smithetal2003} appears to be accurate, for this data, for BAO
models to roughly $\sim5\%$. In light of this, we note that the
discrepancy between {\tt halofit} and the mass power spectra from our
smaller box HR simulations is somewhat puzzling. We highlight this
issue for further study, one possible explanation is the difference in initial 
power spectra, on the other hand, also note that a calculation within the
framework of Renormalized Perturbation Theory
\cite{CrocceScoccimarro2006a} suggests that even for these simulation
volumes on expects small effects due to absence of coupling to large scales.


To fix the small discrepancies which remain, some authors have
advocated making the halo bias factors scale dependent
\cite{Tinkeretal2005}, but the implementation has been based on
fitting formulae rather than fundamental theory.  While
equation~(\ref{eq:NL2Halo}) appears to fare better than the original
approximation (\ref{eq:Lin2Halo}), as we will soon show, in going from
(\ref{eq:Lin2Halo}) to (\ref{eq:NL2Halo}), one is making the
assumption that halo bias is linear even when the mass density field
is not.  If this is not the case then the method is incorrect.  In
addition, there is an unpleasant circularity in requiring prior
knowledge of $P_{\NL}(k)$ in order to predict $P_{\NL}(k)$.


\subsection{Halo bias: The non-linear local bias model}\label{sec:halobias}

The discussion above makes clear that a rigorous treatment of the
2-Halo term is currently lacking.  This term requires a description of
how dark matter haloes cluster.  Whereas current models seek to
describe halo clustering as a biased version of dark matter
clustering, the scale dependence of halo bias is still rather poorly
understood
\cite{ColeKaiser1989,MoWhite1996,MoJingWhite1997,Catelanetal1998,Jing1998,ShethLemson1999,ShethTormen1999,Jing1999,KravtsovKlypin1999,Hamanaetal2001,SeljakWarren2004,Tinkeretal2005}.
In the following sections we develop a model to understand its main
properties.  In particular, we will discuss a general non-linear,
deterministic, local bias model for dark matter haloes.  This model is
exactly analogous to that derived for galaxy biasing by
\cite{FryGaztanaga1993} and first applied to dark matter haloes by
\cite{MoJingWhite1997}.

\def\funk{{\mathcal F}_{\left\{M,R\right\}}}

To begin, consider the density field of all haloes with masses in the
range $M$ to $M+dM$, smoothed with some filter of scale $R$. We now
assume that this field can be related to the underlying dark matter
field, smoothed with the same filter, through some deterministic
mapping and that this mapping should apply independently of the
precise position, $\bx$, in the field: i.e.
\be \delta^{\hc}(\bx|R,M)={\mathcal F}_{\left\{M,R\right\}}
\left[\delta(\bx|R)\right]\ ,\ee
where the sub-scripts on the function ${\mathcal F}$ indicate that it
depends on the mass of the haloes considered and the chosen filter
scale.  The filtered density field is
\be \delta(\bx|R)=\frac{1}{V}\int \dy\, \delta(\by)W(|\bx-\by|,R) \ ,\ee
$W(|x|,R)$ being some normalized filter.  Taylor expanding $\funk$
about the point $\delta=0$ yields
\be \funk\left[\delta(\bx|R)\right]=\sum_i 
\frac{b_i(M|R)}{i!}[\delta(\bx|R)]^i\ .\ee
We now assume that there is a certain filter scale above which $\funk$
is independent of both the scale considered and also the exact shape
of the filter function. Hence,
\be \delta^{\hc}(x|R,M)=\sum_{i=0}^{\infty}\frac{b_i(M)}{i!}
\left[\delta(\bx|R)\right]^{i}\ ,\label{eq:HaloDenFG}\ee
where the bias coefficients are 
\be b_i(M)\equiv\left.\frac{\partial^i\mathcal{F}_{\{M\}}
[\chi]}{\partial\chi^i}
\right|_{\chi=0} \ .\label{eq:biascoeffdeff}\ee
The linear bias model has $b_i = 0$ for all $i>1$.  

The bias coefficients from the Taylor series are not independent, but
obey two constraints. The first arises from the fact that
$\left<\delta^{\hc}(\bx|R)\right>=0$, which leads to
\be b_0=-\frac{b_2}{2}\left<\delta^2\right>-\frac{b_3}{3!}
\left<\delta^3\right>-\dots-\frac{b_n}{n!}\left<\delta^n\right> \ .\ee
Thus, in general $b_0$ is non-vanishing and depends on the hierarchy
of moments. This allows us to re-write equation (\ref{eq:HaloDenFG})
as
\be \delta^{\hc}(x|R,M)=\sum_{i=1}^{\infty}\frac{b_i(M)}{i!}
\left\{\left[\delta(\bx|R)\right]^{i}-\left<\delta^{i}(\bx|R)\right>\right\}\
\label{eq:HaloDenFG2}\ .\ee
Nevertheless, we may remove $b_0$ from further consideration by
transforming to the Fourier domain, where it only contributes to
$\delta(\bk=0)$.

The second constraint states that a sum over all halo density fields
$\delta^{\hc}(x|M)$ weighted by halo mass and abundance must recover
the dark matter density field \cite{Scoccimarroetal2001}.  This
requires that
\be \frac{1}{\rhob}\int dM n(M) Mb_i(M)=\left\{
\begin{array}{l}
1\ \ (i=1)\\
0\ \ (i=0,2,3,\dots)\\
\end{array}
\right.\ .
\label{eq:biasconstrint2} \ee

\begin{figure}
\centerline{
\includegraphics[width=0.8\hsize,angle=-90.0]{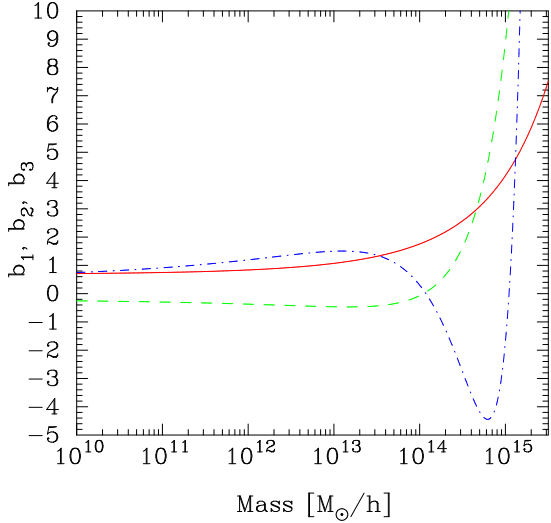}}
\caption{\small{First three halo bias parameters derived from the
Sheth-Tormen \cite{ShethTormen1999} mass function as a function of
halo mass \cite{Scoccimarroetal2001}. The solid line shows 
$b_1$, the dashed line $b_2$, and the dot-dashed line $b_3$.}
\label{fig:halobiasparams}}
\end{figure}

For CDM models whose initial density perturbations are Gaussian
Random, the bias coefficients may either be derived directly through
the `peak--background split' argument
\cite{ColeKaiser1989,MoWhite1996,MoJingWhite1997,ShethTormen1999} or
measured directly from $N$-body simulations.
Figure~\ref{fig:halobiasparams} shows the halo bias parameters up to
third order, derived in the context of the Sheth-Tormen mass function;
see \cite{Scoccimarroetal2001} for the analytic expressions.  We
compare these with measurements from our simulations in
Appendix~\ref{app:MeasureBias}.


A practical application of this method rests squarely upon our ability
to truncate the Taylor series at some particular order.  However,
since the procedure that we have adopted for doing this requires some
further knowledge, we shall reserve our discussion until Section
\ref{ssec:converge}.


\subsection{HaloPT: Evolution of halo fields}\label{sec:HaloPT}

We now evolve the halo density field as expressed by equation
(\ref{eq:HaloDenFG}) into the non-linear regime via perturbation
theory techniques. For a short discussion of these methods see
Appendix \ref{app:PTkernels}, and for a full and detailed review see
\cite{Bernardeauetal2002}. The main idea that we require from
perturbation theory is that each Fourier mode of the density field may
be expanded as a series,
\be \delta(\bk,a)=\sum_i D_1^i(a)\delta_i(\bk) \ ,\ee
where $\delta_i(\bk)$ is the $i$th order Eulerian perturbation and
$D_1(a)$ is the linear growth factor. Thus, on Fourier transforming
the halo bias relation of equation (\ref{eq:HaloDenFG}), truncated at
third order, and on inserting the PT expansion from above, we arrive
at (keeping up to cubic terms)
\def\dq{{d^3\!q}}
\begin{eqnarray}
\delta^{\hc}(\bk|M,R) & = & b_1(M)\left[\frac{}{}\delta_1(\bk|R)+
\delta_2(\bk|R)+\delta_3(\bk|R)\right]\nonumber\\ & &
+\frac{b_2(M)}{2!}\int \frac{\dq_1}{(2\pi)^3}
\left[\frac{}{}\delta_1(\bq_1|R)\delta_1(\bk-\bq_1|R)\right. \nonumber\\
& & \hspace{1.5cm}
\left. \frac{}{}+2\delta_1(\bq|R)\delta_2(\bk-\bq_1|R)\right]\nonumber\\
& & +\frac{b_3(M)}{3!}\int \frac{\dq_1\dq_2}{(2\pi)^6}
\delta_1(\bq_1|R) \delta_1(\bq_2|R)\nonumber\\ 
& & \hspace{1.5cm}\times \frac{}{} \delta_1(\bk-\bq_1-\bq_2|R)\ ,
\label{eq:TaylorPT1} \end{eqnarray}
where $\delta_i(\bq|R) \equiv W(|\bq|R)\delta_i(\bq)$.  We next insert
the solutions for each order of perturbation, which are presented in
equation (\ref{eq:PTdelta}) of the Appendix, into equation
(\ref{eq:TaylorPT1}). On re-arranging terms and collecting powers of
$\delta$, the mildly non-linear density field of dark matter haloes
may be written as a PT series expansion of the dark matter
density. This series is
\ba 
& & \hspace{-0.5cm}
\delta^{\hc}(\bk,a|M,R) =  \sum_{n=1}^{\infty} D_1^{n}(a)
\left[\delta^{\hc}(\bk|M)\right]_n\ 
\label{eq:Halo-PT1a}\ ;
\\
& & \hspace{-0.5cm} 
\left[\delta^{\hc}(\bk|M,R)\right]_n  = \int
\frac{\prod_{i=1}^{n}\left\{d^3\!q_i\,\delta_1(\bq_i)\right\}}{(2\pi)^{3n-3}}
\left[\delta^D(\bk)\right]_n 
\nonumber\\
& & \hspace{2cm}\times 
F^{\hc}_n(\bq_1,...,\bq_n|M,R)
\label{eq:Halo-PT1b}\ea
where $[\delta^{\hc}(\bk|M)]_n$ is the $n$th order perturbation to the
halo density field, and where the short-hand notation
$ \left[\delta^D(\bk)\right]_n=\delta^D(\bk-\bq_1-\dots-\bq_n) $
has been used.  The functions $F^{\hc}_n(\bq_1,...,\bq_n|M,R)$ are the
Halo-PT kernels, symmetrized in all of their arguments. The first
three may be written in terms of the dark matter PT kernels:
\ba 
& & \hspace{-0.2cm}
F^{\hc}_{1} =   \frac{}{}b_1(M)W(|\bk|R) F_{1}
\label{eq:HaloPT-K1}\ ;
\\ 
& & \hspace{-0.2cm}
F^{\hc}_{1,2} =  \frac{}{} b_1(M)W(|\bk|R)F_{1,2}
\nonumber
\\ 
& & \hspace{0.8cm}+ \frac{b_2(M)}{2}W(|\bq_1|R)W(|\bq_2|R) F_1 F_2
\label{eq:HaloPT-K2}
\\
& & \hspace{-0.2cm}
F^{\hc}_{1,2,3} = 
\frac{}{}b_1(M)W(|\bk|R)F_{1,2,3}+ \frac{b_2(M)}{3}
\left[\frac{}{}W(|\bq_1+\bq_2|R)\right. \nonumber 
\\
& & 
\hspace{0.8cm} 
\left. + \frac{}{}W(|\bq_3|R)F_{1,2}F_3 + 2\ \cyc\ \right]\nonumber
\\
& & 
\hspace{0.8cm} +\frac{b_3(M)}{6} W(|\bq_1|R)W(|\bq_2|R)W(|\bq_3|R) F_1
F_2 F_3\ .
\label{eq:HaloPT-K3}\ 
\ea
where $F^{\hc}_{1,\dots,j}\equiv F_j^{\hc}(\bq_1,\dots,\bq_j|M,R)$.
Thus equations (\ref{eq:Halo-PT1a}) and (\ref{eq:Halo-PT1b}) can be
used to describe the mildly non-linear evolution of dark matter halo
density fields to arbitrary order in the dark matter perturbation, and
equations (\ref{eq:HaloPT-K1}--\ref{eq:HaloPT-K3}) make explicit the
halo evolution up to 3rd order.  Together, these ideas define our
meaning of the term `Halo-PT'.

It is now apparent that halo clustering studies which assume a linear
bias model and take the power spectrum to be the fully non-linear one,
are effectively assuming that
$\delta^{\hc}(\bk|M)=b_1(M)\sum\delta_i(\bk)$, with $b_i(M)\ll b_1(M)$
for $i\ne1$.  However, for CDM, the peak-background split argument
informs us that this never happens, unless the density field itself is
linear
\cite{ColeKaiser1989,MoWhite1996,MoJingWhite1997,ShethTormen1999,
Scoccimarroetal2001}.  {\em We must therefore conclude that
extrapolating the linear bias relation into the weakly non-linear
regime, without full consideration of the non-linearity of halo bias
is incorrect. }


\section{The 1-Loop Halo Model}\label{sec:HaloPow}

\subsection{Halo center power spectra}

We now use the Halo-PT to calculate the power spectrum of halo centers
in the mildly non-linear regime.  We define the power spectrum of halo
centers for haloes with masses $M_1$ and $M_2$ to be
\be
\left<\delta^{\hc}_{\cent}(\bk_1|M_1)\delta^{\hc}_{\cent}(\bk_2|M_2)
\right>\!=\!(2\pi)^3\!\delta^D\!(\bk_{12})
P^{\hc\hc}_{\cent}(\bk_1|M_1,M_2)
\label{eq:hcpowReal}\ .\ee
On inserting the Halo-PT solutions for each order of the perturbation
we find that $P^{\hc\hc}_{\cent}(k)$ can be written as the sum of
three terms
\ba
P^{\hc\hc}_{\cent}(k|M_1,M_2) 
& = & P^{\hc\hc}_{\cent,11}(k|M_1,M_2)+ P^{\hc\hc}_{\cent,22}(k|M_1,M_2) 
\nonumber\\
& & +2P^{\hc\hc}_{\cent,13}(k|M_1,M_2)\ ,\label{eq:PowHaloHalo1}\ea
where 
\ba
& & P^{\hc\hc}_{\cent,11}(k|M_1,M_2) = 
\frac{}{}F^{\hc}_1(k|M_1)F^{\hc}_1(k|M_2)P_{11}(k)\ ;
\hspace{1cm}\label{eq:P11HaloHalo}\\
& & P^{\hc\hc}_{\cent,22}(k|M_1,M_2) = 2\int\frac{d^3\!q}{(2\pi)^3}
P_{11}(q) P_{11}(|\bk-\bq|) \nonumber\\
& & \hspace{1.5cm}\times 
\frac{}{}F_2^{\hc}(\bq,\bk-\bq|M_1)F_2^{\hc}(\bq,\bk-\bq|M_2)\ 
;\label{eq:P22HaloHalo}\\
& & P^{\hc\hc}_{\cent,13}(k|M_1,M_2)  = 
6P_{11}(k)\!\int\!\frac{d^3\!q}{(2\pi)^3}
P_{11}(q)\nonumber \\
& & \hspace{1.5cm}\times \ \bar{F}^{\hc\hc}_{13}(\bk,\bq|M_1,M_2)\ .
\label{eq:P13HaloHalo}\ea
Here $P_{11}(k)$ is equivalent to the linear theory power spectrum
and 
\ba & & \bar{F}^{\hc\hc}_{13}(\bk,\bq|M_1,M_2) = \frac{1}{2}
\left[F^{\hc}_1(\bk|M_1)F^{\hc}_3(\bk,\bq_1,-\bq_1|M_2)
\right.\nonumber\\ & & \hspace{1cm}\left.+F^{\hc}_1(\bk|M_2)
F^{\hc}_3(\bk,\bq_1,-\bq_1|M_1)\right]\ .  \ea
When these expressions are averaged over all halo masses, 
weighted by the respective cosmic abundances $M_1n(M_1)$ and
$M_2n(M_2)$, then the constraint equation~(\ref{eq:biasconstrint2})
guarantees that they reduce to the standard `1-loop' expression for 
the PT power spectrum of dark matter \citep{Bernardeauetal2002}:
\ba & & \hspace{-0.7cm} P_{\rm 1Loop}(k) =
P_{11}(k)+P_{22}(k)+P_{13}(k)\ .\label{eq:1Loop} \ea
Strictly speaking the 1-Loop power spectrum refers to $P_{22}+P_{13}$,
we shall break convention and use equation (\ref{eq:1Loop}) to define
what we mean. Explicit details of the 1-Loop expressions may be found
in Appendix \ref{app:1loop}.

The theory may be further developed by directly substituting the
Halo-PT kernels, given by equations
(\ref{eq:HaloPT-K1}-\ref{eq:HaloPT-K3}), into equations
(\ref{eq:P11HaloHalo}--\ref{eq:P13HaloHalo}). A little algebra shows
that
\begin{widetext}
\ba & & \hspace{-0.5cm} 
\frac{P^{\hc\hc}_{\cent}(k|M_1,M_2,R)}{\left|W(kR)\right|^2} = \frac{}{}
b_1(M_1)b_1(M_2)P_{\rm 1Loop}(k) +
\bar{b}_{1,3}(M_1,M_2)\sigma^2(R)P_{11}(k)
\nonumber\\ & & +2\,\bar{b}_{1,2}(M_1,M_2)
\int\frac{\dq}{(2\pi)^3}\frac{W(qR)W(|\bk-\bq|R)}{W(kR)}P_{11}(q)
\left\{\frac{}{}P_{11}(|\bk-\bq|)F_2(\bq,\bk-\bq)+
2P_{11}(k)F_2(\bk,-\bq)\right\}\nonumber \\ & &
+\frac{b_2(M_1)b_2(M_2)}{2}\int \frac{\dq}{(2\pi)^3}
\frac{\left|W(qR)\right|^2\left|W(|\bk-\bq|R)\right|^2}
{\left|W(kR)\right|^2}P_{11}(\bq)
P_{11}(|\bk-\bq|)
\label{eq:HaloHaloPTPow1}\ ,\ea
\end{widetext}
where 
\be \bar{b}_{i,j}(M_1,M_2)
\equiv\frac{1}{2}\left[b_i(M_1)b_j(M_2)+b_j(M_1)b_i(M_2)\right]\ .\ee
Before continuing, we point out and answer an important question that
naturally arises at this junction: How does one compare the filtered
theory with the unfiltered observations?  We forward the proposition
that the unfiltered nonlinear power spectrum can be recovered through
the following simple operation:
\be P(k|R)/|W(kR)|^2=P(k) \ .\ee
This is unquestionably true for an observed non-linear field. It will
therefore also be true for the {\em correct} theoretical model. 
 
This rather lengthy expression may be more readily digested through
the examination of two limiting cases.  But first, notice the
important fact that it is still a separable product of mass dependent
terms and scale dependent terms.

When the two halo masses are identical, then 
\begin{widetext}
\ba
& & 
 \frac{P^{\hc\hc}_{\cent}(k|M_1=M_2,R)}{|W(kR)|^2} = 
 b^2_1(M)\, P_{\rm 1Loop}(k) + b_1(M)b_3(M)\, \sigma^2(R)P_{11}(k)\nonumber\\
& & + b_1(M)b_2(M)\,
 \int\frac{\dq}{(2\pi)^3}\,\frac{W(qR)W(|\bk-\bq|R)}{W(kR)}\,P_{11}(q)
\left\{\frac{}{}
2P_{11}(|\bk-\bq|)F_2(\bq,\bk-\bq)+4P_{11}(k)F_2(\bk,-\bq)
\right\}\nonumber\\
& &  +\frac{b^2_2(M)}{2}
\int \frac{\dq}{(2\pi)^3} \frac{|W(qR)|^2|W(|\bk-\bq|R)|^2}{|W(kR)|^2}
 \,P_{11}(q) P_{11}(|\bk-\bq|)
\label{eq:HaloHaloPTPow2}\ .\ea 
\end{widetext}
This expression is equivalent to evolving the non-linear, local, 
galaxy bias model \cite{FryGaztanaga1993} through Eulerian PT.  
This has been explored by \cite{Heavensetal1998} and \cite{Taruya2000}.

Secondly, consider the case where we integrate over one of the halo
masses, say $M_2$, weighting by $M_2$ and its abundance $n(M_2)$.
Equation~(\ref{eq:biasconstrint2}) again insures that all terms
involving $b_2(M_2)$ and $b_3(M_2)$ vanish, and so the resulting
expression is the 1-Loop correction to the halo center--dark matter
cross power spectrum:
\begin{widetext}
\ba \frac{P^{\delta\hc}_{\cent}(k|M,R)}{|W(kR)|^2} & = &
\left\{ b_1(M)P_{\Loop}(k)+
 \frac{1}{2}b_3(M)\sigma^2(R) P_{11}(k)\right\} +
 b_2(M)\int\frac{\dq}{(2\pi)^3} \frac{W(qR)W(|\bk-\bq|R)}{W(kR)}
 \nonumber \\ & &
\hspace{1cm} \times\ P_{11}(q)
\left[\frac{}{}P_{11}(|\bk-\bq|)F_2(\bq,\bk-\bq)+2P_{11}(k)F_2(\bk,-\bq)\right]
\label{eq:HaloPTPow}
\ .\ea
\end{widetext}%

Inspection of these two limiting cases reveals three remarkable features:
\begin{itemize}
\item Firstly, if the non-linear bias parameters $b_2(M)$ and $b_3(M)$
are non-vanishing then \emph{the bias on large scales is not
$b_1(M)$}.
\item Secondly, the halo-halo spectrum has a term that
corresponds to constant power on very large scales, whereas the cross
spectrum does not. 
\end{itemize}
Both of these points were independently noted by
\cite{Heavensetal1998} and \cite{Taruya2000}, but for the case of
non-linear galaxy biasing (also see \cite{ScherrerWeinberg1998}).
\begin{itemize}
\item Thirdly, the large scale bias derived from the halo-dark 
matter cross power spectrum is not $b_1$, nor is it given by the 
bias derived from the halo-halo power spectrum.
\end{itemize}

To see these points more clearly we take the $k\rightarrow0$ limit of
equations (\ref{eq:HaloHaloPTPow1}), (\ref{eq:HaloHaloPTPow2}) and
(\ref{eq:HaloPTPow}):
\begin{widetext}
\ba 
& & \hspace{-0.5cm}
P^{\hc\hc}_{\cent}(k|M_1,M_2,R) = b_1(M_1)b_1(M_2)P_{11}(k)
\left\{1
+\sigma^2(R)\left[
 \frac{34}{21}\left(\frac{b_2(M_1)}{b_1(M_1)}+\frac{b_2(M_2)}{b_1(M_2)}\right)
+\frac{1}{2}\left(\frac{b_3(M_1)}{b_1(M_1)}+\frac{b_3(M_2)}{b_1(M_2)}\right)
\right]\right\}\nonumber\\
& & \hspace{4cm}
+\frac{b_2(M_1)b_2(M_2)}{2}\int\frac{\dq}{(2\pi)^3}
\left[P_{11}(q)\left|W(qR)\right|^2\right]^2 
\label{eq:HaloHaloPT1:k_0}\ ;
\ea
\ba
& & \hspace{-0.5cm}
P^{\hc\hc}_{\cent}(k|M,R) =  b^2_1(M)P_{11}(k)\left\{1 
+ \sigma^2(R)\left[\frac{68}{21}\frac{b_2(M)}{b_1(M)}
+\frac{b_3(M)}{b_1(M)}\right]\right\}+\frac{b^2_2(M)}{2}
\int\frac{\dq}{(2\pi)^3}\left[P_{11}(q)|W(qR)|^2\right]^2 
\label{eq:HaloHaloPT2:k_0}\ ;\ea
\ba
& & \hspace{-0.5cm}
P^{\delta\hc}_{\cent}(k|M,R) =  b_1(M)P_{11}(k)\left\{1 + \sigma^2(R)
\left[\frac{34}{21}\frac{b_2(M)}{b_1(M)}+\frac{1}{2}
\frac{b_3(M)}{b_1(M)}\right]\right\} \label{eq:HaloPT:k_0} \ .
\ea
\end{widetext}
%


These expressions make the first point noted above trivially obvious:  
The large scale bias is modulated by the Halo-PT correction terms, 
and these depend on the non-linear bias parameters $b_2$ and $b_3$ 
and also on the filtered variance of fluctuations.

The second point noted above originates specifically from the
quadratic non-linear bias terms found in equations
(\ref{eq:HaloHaloPT1:k_0}) and ({\ref{eq:HaloHaloPT2:k_0}), e.g.
terms containing $b_2(M_1)b_2(M_2)$ and $b^2_2(M)$. For a linear power
spectrum that obeys the limit $P_{11}(k\rightarrow 0)\rightarrow0$,
these expressions reduce to the constant
\be P^{\hc\hc}_{\cent}(k\rightarrow0|M)=\frac{b^2_2(M)}{2}
\int\frac{\dq}{(2\pi)^3}\left[P_{11}(q)|W(qR)|^2\right]^2 \ .\ee
This term was discussed in great detail for the case of galaxy biasing
by \cite{Heavensetal1998}.
 
The third point noted above can be understood by constructing the
linear bias, e.g. dividing equations (\ref{eq:HaloPT:k_0}) and
(\ref{eq:HaloHaloPT2:k_0}) by $P_{11}$. On squaring the bias recovered
from (\ref{eq:HaloPT:k_0}) and subtracting it from the bias from
(\ref{eq:HaloHaloPT2:k_0}), we find 
\ba & &
\hspace{-0.4cm}\left\{[b^{\hc\hc}_{\Lin}]^2-[b^{\delta\hc}_{\Lin}]^2\right\} = 
\frac{b^2_2(M)}{2P_{11}(k)}
\int\frac{\dq}{(2\pi)^3}\left[P_{11}(q)|W(qR)|^2\right]^2
\nonumber\\ 
& & \hspace{1.7cm}-\sigma^4(R) \left\{ \frac{34}{21}b_2(M)+\frac{1}{2} 
b_3(M)\right\}^2 
,\label{eq:PhPcross}\ea
We now see that, because $P^{\hc\hc}$ approaches a constant on very
large scales, on dividing through by $P_{11}(k|M)$ the bias function
$b^{\hc\hc}$ diverges at the origin as $1/P_{11}^{1/2}$ diverges.

Figure \ref{fig:LargeScaleLim} shows our expressions for
$b^{\hc\hc}_{\Lin}$ and $b^{\delta\hc}_{\Lin}$ (from equations
\ref{eq:HaloHaloPT1:k_0} and {\ref{eq:HaloHaloPT2:k_0}). In this
particular case, we assume the nonlinear bias parameters derived from
the Sheth-Tormen mass function by \cite{Scoccimarroetal2001}.  We use
a Gaussian filter, for which $W(kR_{\rm G})=\exp[-(kR)^2/2]$.  For
$R_{\rm G}=20 h^{-1}\Mpc$ and for our fiducial cosmology, we find
$\sigma^2(R_{\rm G}=20)\approx0.046$.  To inspect the differences more
closely, we take the ratio of the predictions with respect to the
tree-level theory,
e.g. $b_{1}=P^{\delta\hc}_{11}/P_{11}=\sqrt{P^{\hc\hc}_{11}/P_{11}}$.
The figure demonstrates two of the points raised above. Firstly the
large scale bias is not simply $b_1$: halo-halo bias (solid through to
dotted curves) does not converge as one considers larger and larger
scales; however bias from the cross power spectrum is very close to
linear for all except the most massive haloes, where $b_2$ and $b_3$
are very strongly rising functions (see
Fig. \ref{fig:halobiasparams}).  Secondly, it is now obvious that
$b^{\hc\hc}$ and $b^{\delta\hc}$ are not the same. Note also that the
magnitude of the expected scale dependence: the bias varies by at most
five percent when $k$ is changed by an order of magnitude.  The mass
dependence of $b^{\hc\hc}$ shown in Fig.~\ref{fig:LargeScaleLim} is
simply driven by that of $b^2_2(M)$ (again see
Fig. \ref{fig:halobiasparams}).  As halo mass increases the bias
slowly increases until it reaches a maximum at
$M\sim10^{13}h^{-1}M_{\odot}$. It then decreases to
$M\sim10^{14}h^{-1}M_{\odot}$ after which it shoots up dramatically
for larger masses.


\begin{figure}
\centerline{
\includegraphics[width=8cm]{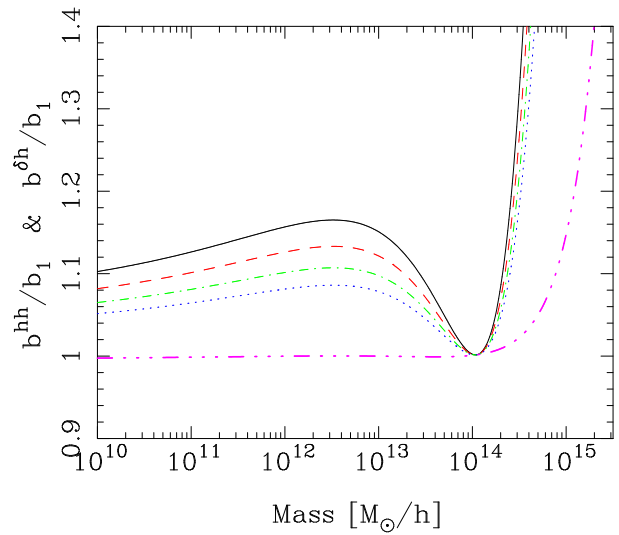}}
\caption{\small{`1-Loop' bias parameters $b^{\hc\hc}$ and $b^{\delta\hc}$ in
the ultra-large scale limit. The solid through to dotted curves show
$b^{\hc\hc}$, measured on scales ($k_{\rm obs}=\left\{0.001,\ 0.005,\
0.01,\ 0.05\right\} h{\rm Mpc}^{-1}$), derived from the 1-Loop
halo-halo cross power spectrum as a function of halo mass.  The
triple-dot dash curve shows $b^{\delta\hc}$ as derived from the 1-Loop
halo-dark matter cross power spectrum.
\label{fig:LargeScaleLim}}}
\end{figure}


\subsection{Convergence of the power spectrum}\label{ssec:converge}

We now return to the issue of truncation and applicability of the
Taylor series expansion of the halo field.  A first requirement for
the Taylor series to converge after a finite number of terms is that
the filter scale be large enough so that the rms dark matter
fluctuations be much less than unity: $\sigma(R)\ll1$. Consider the
case where $R$ is very large and convergence occurs at first order, we
then have: $\delta^{\hc}(\bx|R)=b_1(M)\delta_1(\bx|R)$. As the filter
scale $R$ is slowly decreased the rms fluctuations in $\delta$
increase and a larger and larger number of terms are required to
accurately map the underlying bias function. Finally, as
$\sigma(R)\rightarrow1$ all terms in the series are required. 
At this point the method has no merit.

Since a robust criterion for truncation is out of reach at present, we
propose an {\em ad-hoc} criterion for convergence that must plausibly
be obeyed, that is
\be b_1(M)\sigma(R)<1 \label{eq:TaylorConverge}\ .\ee
In Appendix~\ref{app:MeasureBias} we shall also discuss an empirical
method for testing convergence.


\subsection{Returning to the halo model}\label{ssec:returning}

We now translate these ideas back into the language of the halo
model. To begin, we shall restrict our attention to the halo model in
the large scale limit, more precisely we consider scales where
$U(k|M)\sim1$.  Since the Fourier transform of the mass normalized
profile may be written
\be U(k|M)=\int_{0}^{r_{\rm vir}} \dr\, U(r|M)\, \frac{\sin[kr]}{kr}\ ,\ee
our large scale condition simply becomes $kr_{\rm vir}\ll1$. In the
above equation we have, for convenience, assumed spherical density
profiles. If halo mass and virial radius are related through $M=4\pi
200 \rhob r_{\rm vir}^3/3$, then for the largest collapsed objects in
the Universe $M\approx 10^{15}h^{-1}M_{\odot}$, and the above
condition translates to the inequality $k\ll 0.4 h\Mpc^{-1}$.  If we
assume an NFW density profile \cite{NavarroFrenkWhite1997} with
concentration parameter $c\sim6$, then for $k\sim0.15 h\Mpc^{-1}$, we
find that $U(k)\approx0.994$. We therefore assume that this is an
excellent approximation over the scales that we are interested in.

Hence for scales $kr_{\rm vir}\ll1$, our equations (\ref{eq:Pow1H})
and (\ref{eq:Pow2H}), at the 1-Loop level in Halo-PT, now take the
forms:
\ba 
\frac{P^{ij}_{\1H}(\bk|R)}{\left|W(kR)\right|^2 } & = & \frac{1}
{\rhob^{i}\rhob^{j}}\int_{0}^{\infty} dM n(M) M^2 \nonumber \\
& & \times\Theta_{ij}(M,M)\label{eq:Pow1H1Loop}\ ; \\
\frac{P^{ij}_{\2H}(\bk|R)}{\left|W(kR)\right|^2 } 
& = & \frac{1}{\rhob^i\rhob^j}\int_{0}^{\infty}
\prod_{l=1}^{2} \left\{dM_l n(M_l) M_l\right\} \nonumber
\\ & & \hspace{-0.8cm}
\times\ \frac{P^{\hc\hc}_{\cent}(\bk|M_1,M_2,R)}
{\left|W(kR)\right|^2 }\Theta_{ij}(M_1,M_2)
\ ,\label{eq:Pow2H1Loop} 
\ea
where we have explicitly included a filter on the 1-Halo term.
(Recall that the halo center power spectrum in the 2-Halo term already
includes such a filter.)

We now see that, because the $b_i(M)$ are the only mass dependent
functions, on insertion of equation (\ref{eq:HaloHaloPTPow1}) into
equation (\ref{eq:Pow2H1Loop}) the integrals over mass may be
immediately computed. Thus,
\be P^{ij}_{\2H}(\bk|R)=P^{ij,\hc\hc}(\bk|R) \ ,\ee
where $P^{ij,\hc\hc}(\bk|R)$ is equivalent to equation
(\ref{eq:HaloHaloPTPow1}), except that we have replaced all of the
mass dependent bias parameters by the average ones: i.e.
\be \left<b_i\right>\equiv \frac{\int_{0}^{\infty} dM n(M) M b_i(M)
 \Theta(M-M_{\rm cut})}{\int_{0}^{\infty} dM n(M) M \Theta(M-M_{\rm
cut})}\ .\ee
We note that if we wish to weight by halo number density rather than
mass density then we simply remove the mass weighting in the numerator
and denominator of $\left<b_i\right>$.


\begin{figure*}
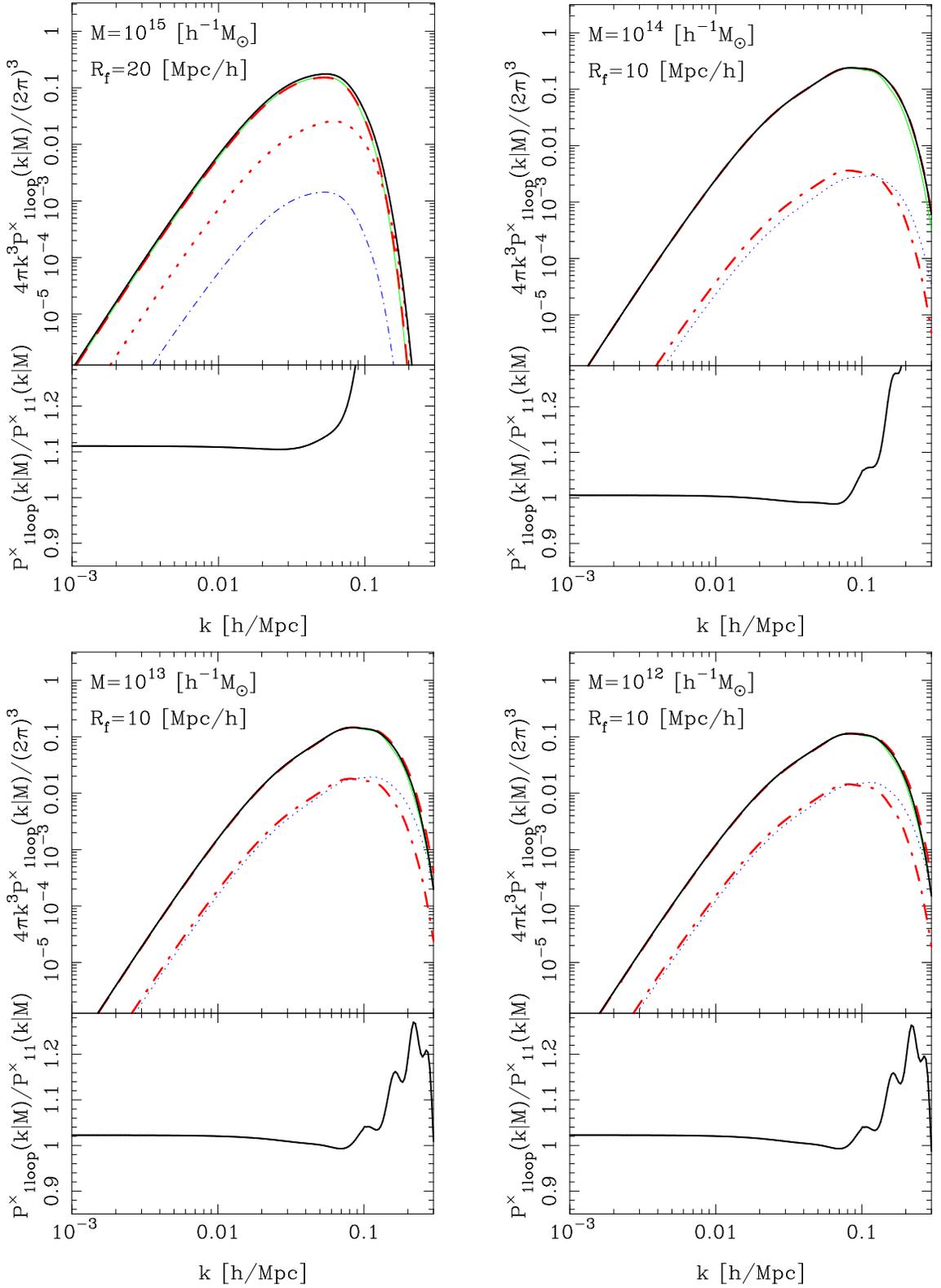

\centerline{
\includegraphics[width=7cm]{Fig.6a.ps}\hspace{1cm}
\includegraphics[width=7cm]{Fig.6b.ps}}\vspace{0.2cm}
\centerline{
\includegraphics[width=7cm]{Fig.6c.ps}\hspace{1cm}
\includegraphics[width=7cm]{Fig.6d.ps}}
\caption{\small{The real space cross power spectrum of haloes and cold
dark matter at the 1-Loop level on large scales. The four panels show
predictions for haloes with masses: $M=\left\{10^{15},\ 10^{14},
10^{13},\ 10^{12}\right\} h^{-1}M_{\odot}$. In the upper plot of each
panel, the thick (black) curve represents the total spectrum as given
by equation (\ref{eq:HaloPTPow}). The dash curve gives the 1-Loop
contribution from the linear bias parameter $b_1$ term; the dotted
curve and dot-dash curves give the 1-Loop contributions from the
non-linear bias parameters $b_2$ and $b_3$, and the thickness (color)
of the lines indicates their sign, with thick (red) lines being
positive and thin (blue) lines being negative contributions.  The thin
solid (green) line gives the smoothed, linearly biased, linear power
spectrum. The lower plot of each panel shows the ratio of the total
cross spectrum with the smoothed linear theory cross spectrum.}
\label{fig:HaloPTPow}}
\end{figure*}


\begin{figure*}
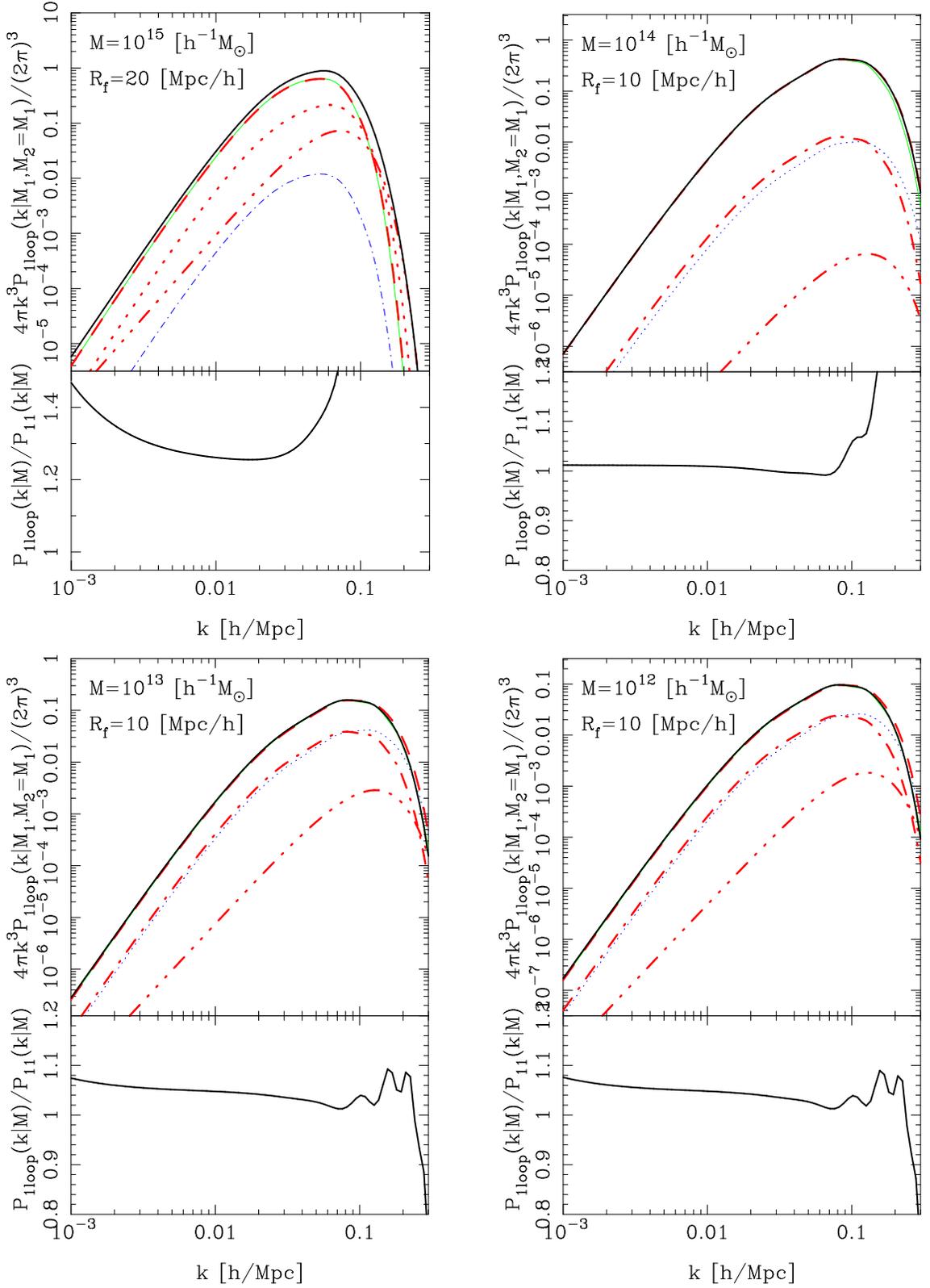

\centerline{
\includegraphics[width=7cm]{Fig.7a.ps}\hspace{1cm}
\includegraphics[width=7cm]{Fig.7b.ps}}\vspace{0.2cm}
\centerline{
\includegraphics[width=7cm]{Fig.7c.ps}\hspace{1cm}
\includegraphics[width=7cm]{Fig.7d.ps}}
\caption{\small{The real space halo-halo power spectrum at the 1-Loop
level on large scales. The four panels show predictions for haloes
with masses: $M=\left\{10^{15},\ 10^{14}, 10^{13},\ 10^{12}\right\}
h^{-1}M_{\odot}$. In the upper panels of each sub-figure, the thick
(black) curve represents the total spectrum as given by equation
(\ref{eq:HaloHaloPTPow2}), the dash curve represents the 1-Loop
contribution from the pure linear bias ($b^2_1$) term, the dotted and
dot-dash curves represent the 1-Loop contributions from the $b_2b_1$
and $b_3b_1$ terms, respectively. The triple dot-dash curve represents
the quadratic nonlinear bias term $b_2^2$. Lastly, the thin solid
(green) line denotes the smoothed, linearly biased linear theory
spectrum. As before, line thickness (color) distinguishes positive and
negative contributions. Bottom panels show the ratio with the linear
spectrum.}
\label{fig:HaloHaloPTPow}}
\end{figure*}


\section{Evaluation of the theory}\label{sec:results}

In this section we present the results from the direct computation of
the 1-Loop halo center expressions for $P^{\hc\hc}_{\cent}$ and
$P^{\delta\hc}_{\cent}$, as given by equations
(\ref{eq:HaloHaloPTPow2}) and (\ref{eq:HaloPTPow}), respectively.
This section is almost entirely pedagogical; we urge those who are
only interested in the direct comparison with the numerical work to
press on to Section \ref{sec:comparison}.

Recall that it is necessary to adopt some filter scale $R$.  We have
studied two choices: $R_{\rm G}=20h^{-1}\Mpc$ and $R_{\rm
G}=10h^{-1}\Mpc$ for which the linear theory, Gaussian filtered,
variances are $\sigma^2(R)=0.046$ and $0.177$, respectively.  The
larger smoothing scale is required for the more massive haloes.


\subsection{Halo--dark matter cross power spectra}

In Figure \ref{fig:HaloPTPow} we show the predictions for the scale
dependence of $P^{\delta\hc}$ at the 1-Loop level, as a function of
wave number. The different panels show results for different halo
masses and smoothing scales.  In all four panels, we see that, as
expected, there is a small (few percent) positive offset from the
linear theory bias value $b_1$. The largest offset occurs for the
cluster mass haloes, but here it may be the case that, owing to the
bias being large, the filter scale that we have adopted for these
objects may still be too small for adequate convergence. We also note
that the offset for the $M=10^{14}h^{-1}M_{\odot}$ haloes is
negligible.  This can be attributed to the fact that $b_2\approx
b_3\approx0$ (see Fig. \ref{fig:halobiasparams} ). For
the lower mass haloes the offsets are roughly $\sim2\%$ in excess of
linear.

Considering the predictions for the highest mass haloes, we see that
the spectrum is scale independent up to $k~\sim0.07 h\Mpc^{-1}$, where
the non-linear amplification from the $P_{22}$ term becomes
dominant. For this case, the absence of the pre-virialization feature
may be understood as follows: Firstly, we note that $b_1$ and $b_2$
are positive, whereas $b_3$ is negative. However, since $b_3$ is an
order of magnitude smaller than the others, it plays no significant
part in determining the shape of the spectrum. The $P_{13}$ term in
the 1-Loop power spectrum, which is the main cause of the
pre-virialization feature, is thus overwhelmed by the action of the
quadratic bias $b_2$ term. This suggests that $P^{\delta\hc}$ should
be scale independent up to $k~\sim0.07 h\Mpc^{-1}$.

Next, we collectively consider the predictions for the lower mass
haloes, as they show many similar traits. Firstly, we find that when
$k<0.01 h\Mpc^{-1}$, then the ratios of the 1-Loop to tree level (or
linear) spectra are flat. However, for $0.01
h\Mpc^{-1}<k<0.07h\Mpc^{-1}$, significant scale dependence is
apparent: the pre-virialization feature is present and it appears to
become stronger as halo mass decreases. On smaller scales still, the
non-linear boost from the $P_{22}$ term amplifies the power spectrum
and breaks all scale independence. Interestingly, the onset of
$P_{22}$ is pushed to smaller scales as halo mass decreases. These
effects can be understood as follows. For these objects $b_1$ and
$b_3$ are positive, whereas $b_2$ is negative. On large scales, we see
that $b_2$ and $b_3$ are nearly equivalent, but $b_3$ is slightly
dominant, and this results in a small positive correction. Whereas on
smaller scales this trend reverses and $b_2$ becomes dominant. The
overall correction is then negative and this leads to the enhanced
pre-virialization feature and delay of the onset of $P_{22}$.

It is also interesting to note the imprint of the BAO features in the
ratios of the power spectra. The strength of the signal appears to
depend on halo mass and increases as halo mass decreases. As we
discussed in Section \ref{ssec:scaledebendence}, this can be
attributed to the fact that non-linear evolution suppresses BAOs on
small scales, and on taking the ratio with a linear theory spectrum,
we are artificially introducing oscillations.


\begin{figure*}
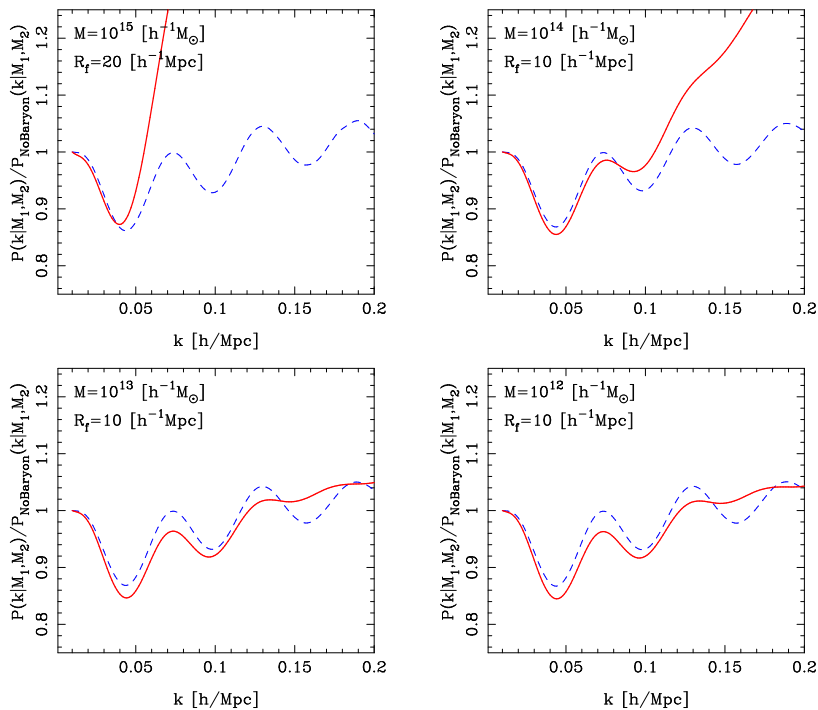

\centerline{
\includegraphics[width=5cm]{Fig.8a.ps}
\hspace{0.5cm}
\includegraphics[width=5cm]{Fig.8b.ps}}
\vspace{0.2cm}
\centerline{
\includegraphics[width=5cm]{Fig.8c.ps}
\hspace{0.5cm}
\includegraphics[width=5cm]{Fig.8d.ps}}
\caption{\small{Non-Linear evolution of the BAOs as traced by the
halo-halo power spectrum for four different halo masses. The solid
(red) line denotes the the non-linear (1-Loop) spectrum as given by
equation (\ref{eq:HaloHaloPTPow2}); the dash (blue) line denotes the
linear theory (tree level) power spectrum. We have taken the ratio of
each spectrum with a smooth `No Baryon' model as described in the
text, and have renormalized each spectrum so that they are unity on a
scale $k\sim0.01 h\Mpc^{-1}$.}
\label{fig:BaryonHaloHaloPTPow}}
\end{figure*}


\subsection{Halo-Halo power spectra}\label{ssec:halohalopower}

In Figure \ref{fig:HaloHaloPTPow} we show the predictions for the
scale dependence of $P^{\hc\hc}$, at the 1-Loop level, as a function
of wavenumber. The four panels show results for haloes with masses in
the same range as Figure \ref{fig:HaloPTPow}.  Again, upper panels
show the contributions from each of the Halo-PT terms in equation
(\ref{eq:HaloHaloPTPow2}), and where the sign of each contribution is
distinguished through line thickness/color.  Sub-panels are as
before.

Some obvious similarities exist between the auto-halo spectra and the
cross spectra. In particular: the large scale bias is not given by
$b_1$; the highest mass halo spectrum shows no sign of the
pre-virialization power decrement; the non-linear boost occurs
increasingly at smaller scales as halo mass decreases; with the
exception of the highest mass haloes, the ratio of the 1-Loop spectra
to the tree level spectra have BAOs imprinted. These effects may all
be understood through the explanations from the previous sub-section.

We also notice some important differences between $P^{\hc\hc}$ and
$P^{\delta\hc}$. Firstly, the addition of the quadratic non-linear
bias term, $[b_2(M)]^2$, modifies the results on the largest
scales. As was discussed in Section \ref{sec:HaloPow},
$P^{\hc\hc}_{\rm 1Loop}$ becomes a white-noise power spectrum as
$P_{11}\rightarrow0$, unless $b_2=0$.  In the figure, the
contributions from this term are denoted by the triple dot dash
lines. On considering all four panels and paying special attention to
the ratios, we see that, with the exception of the case
$M\sim10^{14}h^{-1}M_{\odot}$, there is an upturn in power on scales
$k<0.01 h\Mpc^{-1}$. For the $M\sim10^{14}h^{-1}M_{\odot}$ haloes this
effect is not found, this owes to the fact that $b_2\sim0$ (see Fig
\ref{fig:halobiasparams}).

Secondly, we note that on smaller scales the effect of the $b_2^2$
term is to boost the power across all scales. Thus the
pre-virialization power decrement is no longer a decrement relative to
the linear theory. However, relative to the power measured on say
$k\sim0.01 h\Mpc^{-1}$, there is a very real decrement. Moreover,
the $k$-dependence of the terms multiplying $b^2_2$ will make the
decrement appear larger than we would expect from the non-linear
matter spectrum. 


\subsection{Non-linear evolution of BAOs}
\label{sec:BAO}

We now briefly consider how mode-mode coupling and non-linear biasing
affect the evolution of the BAOs. Figure \ref{fig:BaryonHaloHaloPTPow}
compares the 1-Loop predictions for $P^{\hc\hc}$ with the linear
theory predictions. We again consider the set of halo masses:
$M=\left\{10^{15},\, 10^{14},\,
10^{13},\,10^{12}\right\}h^{-1}M_{\odot}$, and, to emphasize the
evolution of the BAO features, we have taken the ratio of each
spectrum with the `No Baryon' model of equation
(\ref{eq:NoBaryon}). In addition, because we are purely interested in
the scale dependence, we re-normalize all ratios to unity at
$k\sim0.01 h\Mpc^{-1}$.

The top left panel shows results for the highest mass haloes: all but
the first of the BAO troughs has been removed through non-linear
evolution. This is a much more aggressive non-linear evolution than
one expects from considerations of the dark matter alone. Secondly,
the trough appears to have been displaced towards lower
frequencies. These effects arise because both the quadratic bias
$(b_2^2)$ and $b_2b_1$ terms are positive, and so dominate over the
negative $b_3$ and $P^{13}$ terms.  This fact, coupled with the
injection of power from $P_{22}$, means that the non-linear boost
occurs at $k\sim0.05h\Mpc^{-1}$.  Thus, the overall effect on the halo
spectrum is to shift the first trough to smaller $k$.

All the spectra of the lower mass haloes display a pre-virialization
feature, the strength of which increases as halo mass decreases.  In
addition, the number of peaks and troughs which remain in the evolved
spectra increases as halo mass decreases, because the strong
non-linear amplification from the $P_{22}$ term is delayed by the
negative $b_2b_1$ corrections. However, in contrast to the high mass
haloes, for the low mass haloes the first acoustic trough is shifted
towards larger $k$ because, when $0.01<k<0.05$, then the negative
correction from the $b_2b_1$ term is dominant and so subtracts power
from the higher frequency side of the oscillation.  This acts to shift
the overall pattern to higher frequencies.


\section{Comparison with simulations}\label{sec:comparison}

In this Section, we compare the predictions from the theoretical model
with the results from the numerical simulations presented in Section
\ref{sec:N-Body}. 


\subsection{Halo center power spectra re-visited}

Returning to our analysis of Figure \ref{fig:romansimHaloPT}, We are
now in a position to comment on the success of the Halo-PT model in
comparison to the numerical simulations. The analytic model was
evaluated using the semi-empirical bias parameters that were
determined as described in Appendix \ref{app:MeasureBias}.  Since here
we are concerned purely with the scale-dependence of the spectra and
not their overall amplitude, we allowed the very large scale
normalizations to be considered as free parameters.  These were
fit-for in exactly the same way as was done for the linear theory
models in equation (\ref{eq:amplitude}). Whilst this re-scaling would
not be necessary if the bias parameters that we were adopting were
precise and accurate, since we have not been able to establish this in
a robust manner we feel that this approach is acceptable, given that
it has been equally applied to the linear theory.  Moreover, this is
the method of analyzing real data.

These predictions are shown as the solid lines in
Figure~\ref{fig:romansimHaloPT}.  Comparison with the measured
$P^{\hc\hc}_{\cent}$, shows that the model and the shot-noise
corrected data show reasonable correspondence over scales
$k<0.07h^{-1}\Mpc$. On smaller scales the match is poor.  However, the
uncertainty in the shot-noise correction makes it very difficult to
draw firm conclusions.

Considering $P^{\delta\hc}$, we find that the analytic model is in
good agreement with the simulations over a much larger range of
scales.  The model correctly captures the mass dependence of the
pre-virialization feature -- low mass haloes show greater loss of
memory to the initial density fluctuations.  The shifting of the
nonlinear boost to smaller scales as halo mass decreases is also
matched rather well.

To assess whether $P^{\delta\hc}_{\cent,\rm 1Loop}$ is a better fit to
the simulation data than is linear theory, we have performed a
likelihood ratio test, assuming that the likelihood functions are
Gaussian.  In this case, a necessary statistic for model selection is
for
\be {\rm LR}\equiv\frac{{\mathcal L}
(\left\{x_l,y_l\right\}|[P^{ij}_{\rm 1Loop}]_{\rm max})} {{\mathcal
L}(\left\{x_l,y_l\right\}|[P^{ij}_{\Lin}]_{\rm max})} >1\ .  \ee
where the subscript `max' refers to the parameter choices in the
models that maximize the likelihood. Restricting the information to be
$k<0.1h\Mpc^{-1}$, we find, going from low to high mass haloes, that
${\rm LR}=\left\{1.06,1.03,1.05,1.07\right\}$, respectively.


\subsection{Halo bias re-visited}

We now examine how well the Halo-PT model does at matching the
nonlinear scale dependent bias of the halo centers as measured in the
numerical simulations (Figure \ref{fig:romansimbias}).  The top
section of each panel shows that the analytic model (solid blue and
red lines correspond to $b^{\delta\hc}_{\rm NL}$ and $b^{\hc\hc}_{\rm
NL}$, respectively) captures, qualitatively, the scale dependence of
the bias.  The model shows a bias that increases with $k$ for the high
mass haloes but decreases with $k$ for lower masses. However, there
are some notable discrepancies: The model under predicts and then over
shoots the measured relationship for the most massive haloes; for the
next bin in halo mass (top-right panel), the measured bias is flat,
whereas the model predicts a down turn after $k\sim0.1h\Mpc^{-1}$; the
model fares better in the two lowest mass bins, but the down-turn at
high $k$ is not seen in the data.  However, we must stress that, with
the exception of the $M\sim10^{14}h^{-1}M_{\odot}$ haloes, the model
out-performs linear theory, which would predict constant bias on all
scales.

Having extolled the virtues of our model we now draw attention to its
short-comings. Whilst the predictions provide a good match to the halo
power spectra, they do not simultaneously provide a good match to the
scale dependence of the bias. If we re-consider our measurements of
the non-linear matter power spectrum (upper sections in Fig
\ref{fig:romansimHaloPT}) $P^{\delta\delta}$, we see that the 1-Loop
model (dot-dash lines) over predicts the LR simulations on scales
$k>0.05 h\Mpc^{-1}$ and the HR simulations on scales
$k>0.07h\Mpc^{-1}$. It is therefore unlikely that the model as
presented here can be made to work precisely .


\section{Galaxy power spectrum}\label{sec:galaxies}


\begin{figure*}
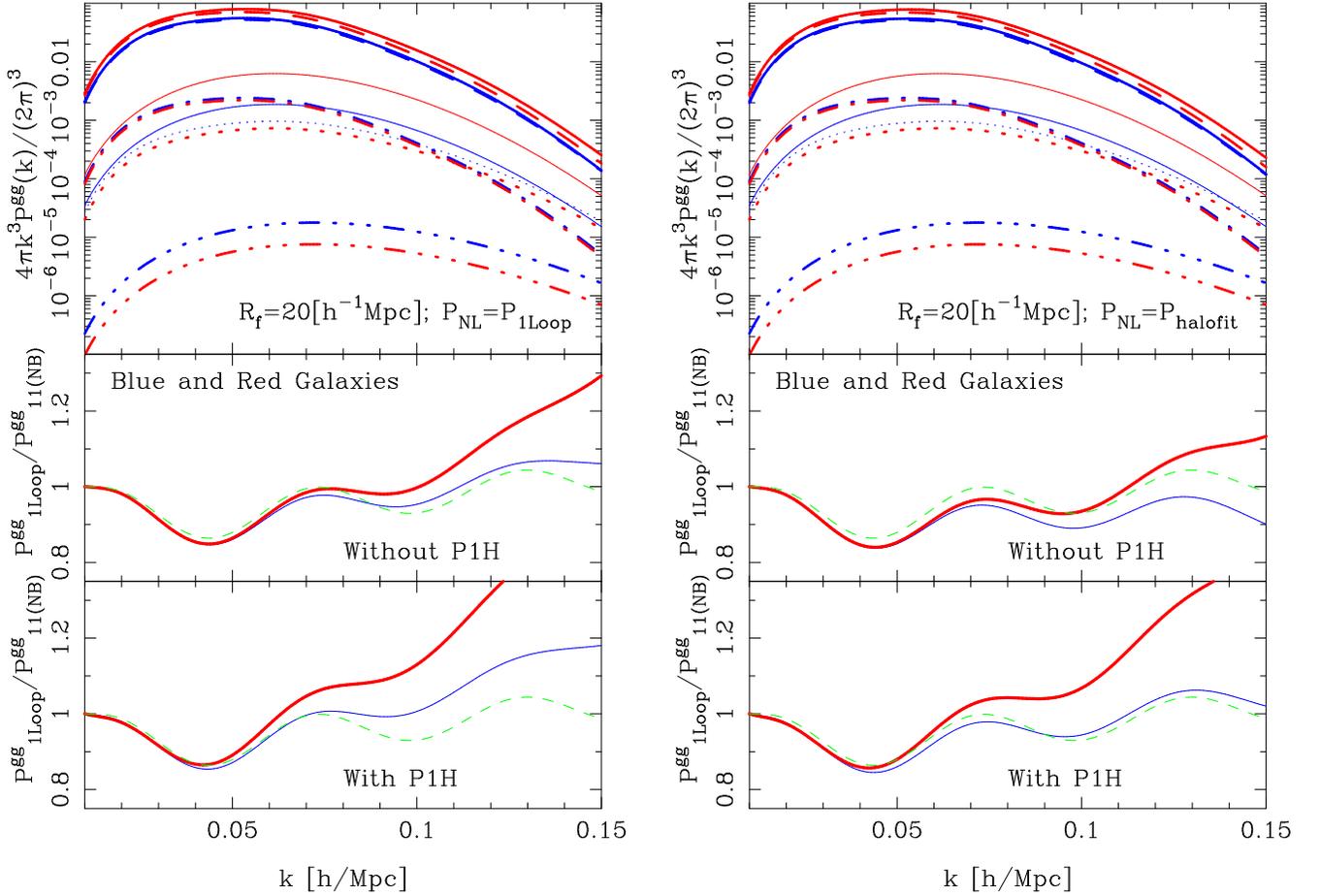

\centerline{
\includegraphics[width=8.5cm]{Fig.9a.ps}\hspace{0.5cm}
\includegraphics[width=8.5cm]{Fig.9b.ps}}
\caption{\small{Scale dependence of the power spectrum of blue and red
galaxies (thick solid blue and red lines) on very large scales. Top
left panel shows the results obtained from the evaluation of equations
(\ref{eq:gal1Halo}) and (\ref{eq:gal2Halo}). The 1-Halo term is
represented by the thin solid lines. For the 2-Halo term we have the
following contributions: $b_1^2$ terms denoted by dash lines; $b_2b_1$
terms denoted by dotted lines; $b_3b_1$ terms denoted by dot-dash
lines; $b_2^2$ terms denoted by triple-dot-dash lines. Note again for
the 2-Halo contributions, line thickness dictates sign. The middle
panel shows the ratio of the red and blue galaxy 2-Halo terms (thick
red and thin blue lines, respectively) with the no-Baryon linear model
of equation (\ref{eq:NoBaryon}). The dash line shows the linear theory
model with BAOs. The bottom panel shows the same as the middle only
this time we show the effect of including the 1-Halo term. The right
hand panel is similar to the left, only we exchange $P_{\rm 1Loop}$ in
the 2-Halo term for the {\tt halofit} model \cite{Smithetal2003}.}
\label{fig:GalaxyPower}}
\end{figure*}


How does the scale dependence of the bias depend on galaxy type?  The
answer has important consequences for future galaxy surveys that will
measure the clustering of specific sub-classes of objects.  We may
address this question using the halo model by changing the mass
weighting in the integrals to:
\ba n(M)M & \rightarrow & n(M)\left<N|M\right> \ ;\nonumber \\ n(M)M^2
& \rightarrow & n(M)\left<N(N-1)|M\right> \ , \ea
where $\left<N|M\right>$ and $\left<N(N-1)|M\right>$ are the first two
factorial moments of the halo occupation probability function
$P(N|M)$, which gives the probability for a halo of mass $M$ to host
$N$ galaxies. Use of the factorial moments of $P(N|M)$ subtracts off a
term corresponding to the self-correlation of galaxies
e.g. $\xi(r\rightarrow0)=\delta^{D}(r)/\nbar$.  This corresponds to
the Poisson shot noise term in Fourier space.  Secondly, the mean
density profile of dark matter is changed to the mean density profile
of galaxies in the halo: $U(k|M)\rightarrow U^{\g}(k|M)$.
Following the discussion in Section \ref{ssec:returning}, it is a very
good approximation to set $U^{\g}(k|M)=1$ when $kr_{\rm vir}\ll 1$.
Thirdly, the constant pre-factors transform as $1/\rhob^2\rightarrow
1/\nbarg^2$, where
\be \nbarg=\int dM n(M)\left< N|M \right> \Theta(M-M_{\rm cut})\ .\ee
And finally, the galaxy bias parameters are
\be b^{\g}_i=\frac{1}{\nbarg}\int dM n(M)\left<N(M)\right> b_i(M)\
.\label{eq:galbiasparam}\ee
These changes in equations~(\ref{eq:Pow1H1Loop})
and~(\ref{eq:Pow2H1Loop}) yield the 1-Loop halo-model prediction for
the galaxy power spectrum. Explicitly:
\ba \frac{P_{\1H}^{\g\g}(k|R)}{|W(kR)|^2 } & = & \frac{1}{\nbar^2}\int dM n(M)
\left<N(N-1)|M\right> \nonumber \\ & & \times \Theta(M-M_{\rm cut})
\ ; \label{eq:gal1Halo}\\
\frac{P_{\2H}^{\g\g}(k|R)}{|W(kR)|^2} & = & 
\frac{1}{\nbar^2}\int \prod_{i=1}^{2}
\left\{\frac{}{}dM_i n(M_i) \left<N|M_i\right> \right. \nonumber\\
& & \times \left. \Theta(M_i-M_{\rm cut})\frac{}{}\right\} \nonumber \\
& & \times \frac{P^{\hc\hc}_{\cent,\rm 1Loop}(k|M_1,M_2,R)}{|W(kR)|^2}\ ,\ea
where again we have explicitly included a filter function on the
1-Halo term. On inserting our expression for $P^{\hc\hc}_{\cent}$ from
equation (\ref{eq:HaloHaloPTPow2}) into the 2-Halo term, and again
noticing that in the large scale limit the mass integrals may be
performed directly, we find:
\begin{widetext}
\ba \frac{P^{\g\g}_{\2H}}{\left|W(kR)\right|^2} & = &
\left\{\frac{}{}[b^{\g}_1]^2P_{1\rm Loop}(k)+ b_1^{\g}\,
b_3^{\g}\, \sigma^2(R) P_{11}(k)\right\}
+\frac{[b_2^{\g}]^2}{2} \int \frac{\dq}{(2\pi)^3} P_{11}(q)
P_{11}(|\bk-\bq|)\frac{|W(qR)|^2 |W(|\bk-\bq|R)|^2}{\left|W(kR)\right|^2}
\nonumber \\
& & \hspace{0cm}+ 2\, b^{\g}_1\,b^{\g}_2\, 
\int\frac{\dq}{(2\pi)^3}P_{11}(q)\frac{W(qR)W(|\bk-\bq|R)}{W(kR)} 
\left\{\frac{}{}
P_{11}(|\bk-\bq|)F_2(\bq,\bk-\bq)+2P_{11}(k)F_2(\bk,-\bq) \right\}
\label{eq:gal2Halo}\ ,\ea
\end{widetext} 
The resulting expression for the galaxy power spectrum is identical to
that given by \cite{Heavensetal1998,Taruya2000}, with one important
difference---we include the large scale constant power originating
from $P^{\g\g}_{\1H}(k|R)$.

To study the expected differences between red and blue galaxies, we
use the parametric forms for $\left<N(M)\right>$ measured by
\cite{ShethDiaferio2001} in the semi-analytic models of
\cite{Kauffmannetal1999}:
\be \left<N_B|M\right>=0.7\left(\frac{M}{M_{B}}\right)^{\alpha_B}\ ;\
\ \left<N_R|M\right>=\left(\frac{M}{M_{R}}\right)^{\alpha_R}\ .
\label{NgalM}
\ee
The blue galaxy parameters are: $M_{B}=4\times10^{12}h^{-1}M_{\odot}$;
for haloes with masses in the range $10^{11}\le M/h^{-1}M_{\odot} \le
4.0\times10^{12}$ then $\alpha_B=0.0$, for larger mass haloes
$\alpha_B=0.8$. The red galaxy parameters are:
$M_{R}=2.5\times10^{12}h^{-1}M_{\odot}$; for haloes with masses
greater than the cut-off mass $\alpha_R=0.9$.  For the second moment
of the HOD we follow the model of Kravtsov et
al. \cite{Kravtsovetal2004}, so that $\left< N(N-1)|M\right> = \left<
N|M\right>^2-1$.  This makes $P(N|M)$ sub-Poissonian as suggested from
the observations
\cite{PeacockSmith2000,YangMovandenBosch2003,Zehavietal2005} and the
semi-analytic models
\cite{Bensonetal2000,ShethDiaferio2001,Scoccimarroetal2001,
Berlindetal2003,Zhengetal2005}; and secondly, this choice allows the
first moment alone to fully specify the hierarchy of moments.

In practice, we use the models above but impose a lower mass cutoff of
$M_{\rm min}=10^{12}h^{-1}M_{\odot}$.  This yields
$\nbar_{B}=4.10\times10^{-3} h^{3}\Mpc^{-3}$,
$\nbar_{R}=7.93\times10^{-3} h^{3}\Mpc^{-3}$, $\left\{b_1^{B}=1.20,\,
b_2^{B}=-0.14,\,b_3^{B}=1.06\right\}$ and
$\left\{b_1^{R}=1.39,\,b_2^{R}=0.09,\,b_3^{R}=0.89\right\}$.  (This
estimate of $b^{B}_2$ agrees with the observational determinations of
similar galaxies from the PSCz survey \cite{Feldmanetal2001}.)  These
values are easily understood by noting how $b_i(M)$ depends on halo
mass (e.g. Figure~\ref{fig:halobiasparams}), the weightings given in
equation~(\ref{NgalM}), and recalling that the halo mass function
declines exponentially at $M>10^{13}h^{-1}M_\odot$.  Notice that the
red galaxy bias parameters are all positive, whereas $b^{B}_2$ for the
blue galaxies is negative.  Therefore, while we expect to see a
pre-virialization feature in $P_B(k)$, we do not for the red galaxies.

The left panel of Figure \ref{fig:GalaxyPower} shows the power
spectrum of blue and red galaxies evaluated using the 1-Loop Halo
model. 
The top section shows the individual contributions 
from all the linear and nonlinear terms.  Note that, for the blue 
galaxies the non-linear correction terms $b_1^{B}b^{B}_2$
(blue dotted lines) and $b_1^{B}b^{B}_3$ (blue dot-dash lines) are
roughly the same order of magnitude as the 1-Halo term (thin blue
line). However, for the red galaxies, the 1-Halo term (thin red line) 
dominates over the non-linear bias corrections by factors of a few. 
For both populations the quadratic bias terms $[b^{\g}_2]^{2}$ 
(triple-dot-dash lines) appear to be negligible. 

The middle section of the left hand panel shows the ratios of the
1-Loop $P^{\g\g}_{\2H}$ with the No-Baryon linear model (from equation
\ref{eq:NoBaryon}).  Again, since we are primarily interested in the
scale dependence of the bias, we re-normalize all curves to be unity
at $k\sim0.01 h\Mpc^{-1}$. As expected, the red galaxy power spectrum
appears to trace the linear theory matter fluctuations (green dash
line) very well when $k<0.07h\Mpc^{-1}$.  The non-linear boost breaks
this accordance at larger $k$.  However, for the blue galaxies, the
scale dependence is more complicated, having an increased
pre-virialization feature and a delayed nonlinear boost.  We also note
that, for the red galaxies, the second and third BAOs have been almost
completely suppressed, whereas only the third peak has been removed
for the blue galaxies.  The peaks and troughs, however, appear to be
in the right places.

So far, we have neglected the contribution from the constant power 
1-Halo term. In the bottom section of the panel we now take this 
into account and show $P^{\g\g}_{\1H}+P^{\g\g}_{\2H}$ ratioed to 
the No-Baryon model.   
For the red galaxies, the agreement between the predictions and the
linear theory that was noticed before is now broken on much larger 
scales, $k\sim0.04h\Mpc^{-1}$. The trough of the first BAO has been 
shifted slightly to smaller $k$ and the non-linear boost occurs
at a larger scale.  Because the 1-Halo term is about 5 times smaller 
for the blue galaxies, the modifications are not as severe. 
The addition of this term offsets the suppression of power caused by 
the negative $b^{B}_1b^{B}_2$ term and the blue galaxies now appear 
to trace the linear theory on scales $k<0.07h\Mpc^{-1}$ quite well. 
At larger $k$ the linear spectrum is a poor match to the 
predictions. We also note that the BAOs are further suppressed and 
the second trough has been shifted to lower frequencies.

Because $P_{\rm 1Loop}$ does not provide a very accurate model for the
true non-linear power spectrum, we have studied the effect of
exchanging $P_{\rm 1Loop}$ for the {\tt halofit} \cite{Smithetal2003}
power spectrum.  The results are shown in the right hand panel of
Figure \ref{fig:GalaxyPower}. Although the predictions are
qualitatively very similar, {\tt halofit} predicts enhanced
pre-virialization and smaller non-linear boosts.  In the middle panel,
where the 1-Halo term is not included, we see that the red and blue
galaxies do not match the linear theory as well on large scales. In
particular the blue galaxy power is suppressed on all scales except
the largest. We also see that the BAOs have been better
preserved. Although there are slight shifts in the positions of the
second trough and third peak.  However, the bottom panel shows that
once the 1-Halo term has been included, the red galaxy predictions are
almost as before. The blue galaxies still show a reasonably strong
pre-virialization feature, but, because of the weak nonlinear boost,
they match the linear theory rather well over nearly all the scales
considered. 

 We note that this modification is not entirely self-consistent
and as such is not meant to be blindly trusted, since the non-linear
bias terms are still derived from the 1-Loop Halo-PT. However, we use
this operation to highlight that a more advanced understanding of the
non-linear power spectrum does change the results quantitatively.
This implies that a more advanced model of the scale dependence of the
bias will also further modify and improve the predictions.


\section{Scale dependent bias at $z>0$}


We have shown that the BAO harmonic series in the power spectrum
at $(z=0)$ is affected by nonlinear effects from bias and
gravitational evolution.  Although naively one might expect that at
higher redshift nonlinear effects are less important, this is not
necessarily so. First, one must pick a criterion for how to compare
things at different redshifts. A natural choice is to use objects of
the same number density. In this case, as it is well-known, at higher
redshifts objects of the same number density are more biased, leading
to stronger nonlinear bias effects, even though the dark matter has
less nonlinear evolution. Therefore, overall it is not clear {\em a
priori} whether the situation improves or not.  Figure
\ref{fig:HaloPower_z1z2} shows the halo power spectra at $z=1$ and
$z=2$ measured from our 8 LR simulations, where the halo samples were
harvested so that they would have the same fixed comoving number
density as the Bin 1 sample at $z=0$ (see from
Table~\ref{table:halocat}). The power spectra analysis was identical
to that as described in Section \ref{sec:N-Body}. This figure clearly
demonstrates that the non-linear bias effects that are present at
$z=0$ (Fig. \ref{fig:romansimHaloPT}), remain present in the high
redshift halo samples. In light of this, we anticipate that low mass
halo samples at higher redshift, constructed so that $M<M^*(z>0)$,
will likewise show enhanced pre-virialization ($M^{*}(z)$ is defined
to be the halo mass at which $\sigma(M)=1$).  We reserve further
details of this issue for future work.


\begin{figure*}
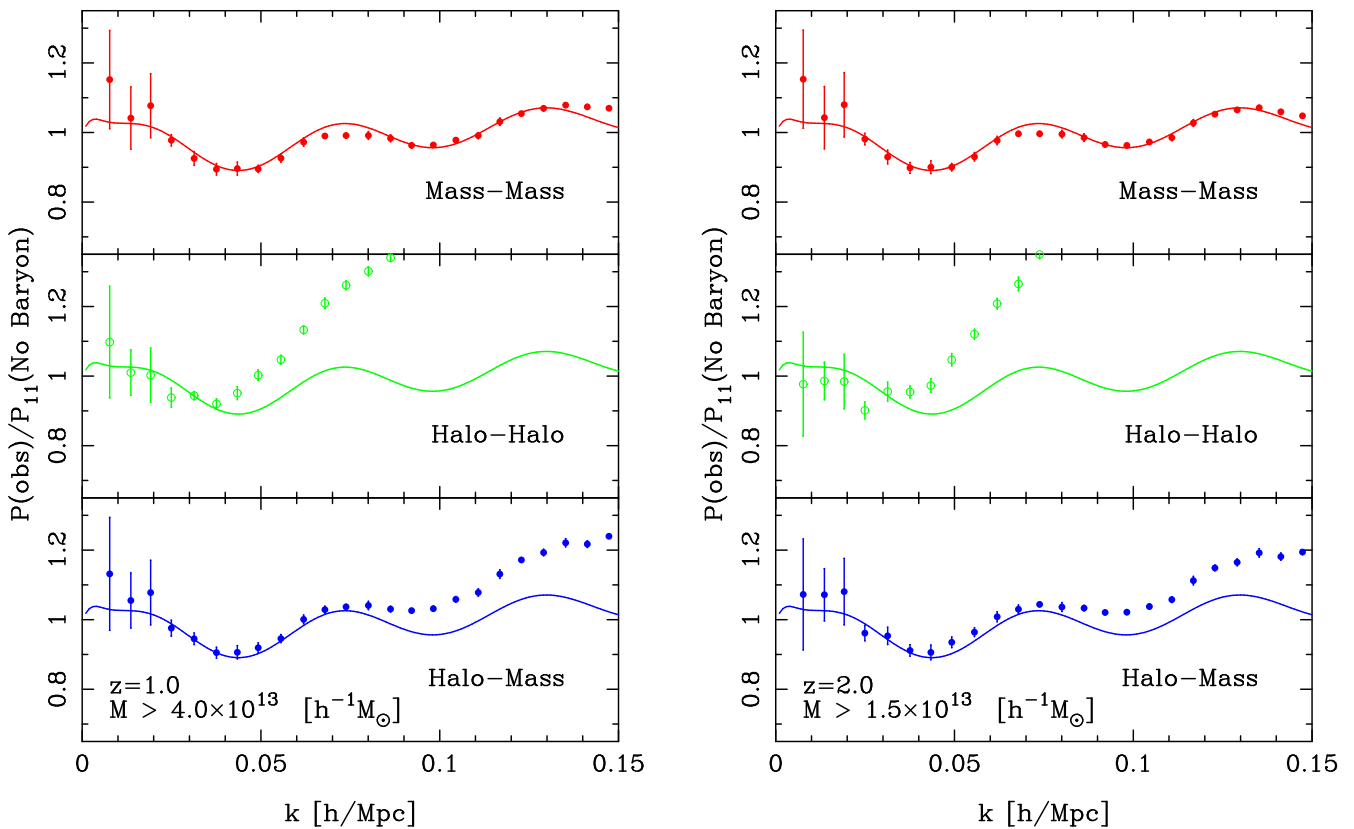

\centerline{
\includegraphics[width=8.5cm]{Fig.10a.ps}
\hspace{0.5cm}
\includegraphics[width=8.5cm]{Fig.10b.ps}}
\caption{\small{Dark matter, halo--halo and halo--dark matter cross
power spectra at $z=1$ (left panel) and $z=2$ (right panel) as a
function of wavenumber. Symbol styles are as in Fig
\ref{fig:romansimHaloPT}. At both epochs we computed the halo spectra
using the same fixed number density of haloes as those in Bin 1 sample
from $(z=0)$: $\bar{n}=3.43\times10^{-5}[h/{\rm Mpc}]^3$.  Non-linear
mass and bias corrections remain present at the level of 5-10\% in the
halo spectra.}
\label{fig:HaloPower_z1z2}}
\end{figure*}


\section{Conclusions}\label{sec:conclusions}

In this paper we have explored in detail, through both numerical and
analytic means, the scale dependence of the nonlinear dark matter,
halo and galaxy power spectra on very large scales $k<0.15
h\Mpc^{-1}$.  For our numerical work we used and ensemble of 20
simulations in boxes of side $512h^{-1}\Mpc$ and 8 simulations in
boxes of side $\sim1h^{-1}$Gpc.  Each simulation contained more than
134 million particles. We have found that:
\begin{itemize}
\item 
The non-linear matter power spectrum is suppressed relative to the
linear theory by $5-10\%$ on scales between
$0.05<k/(h\Mpc^{-1})<0.075$ at the 2-$\sigma$ level.
\item 
The halo-dark matter cross power spectrum shows clearly that the bias
of halo centers is non-linear on very large scales.  The form of the
non-linearity depends strongly on halo mass: for high mass haloes no
pre-virialization suppression is seen; whereas there is an apparent
$\sim10\%$ suppression of power relative to linear theory for lower
mass haloes.
\item
To make robust statements concerning their large scale clustering of
haloes, it is essential to characterize the shot noise correction to
high precision.  For high mass haloes this correction is
sub-Poissonian, so the simple and widely used $1/\nbar$ model must be
inappropriate. Halo exclusion effects lead to a plausible explanation
for this phenomenon, which we used to motivate an alternate
correction. However further work is required to establish this
robustly. In addition the true answer may need to take into account
the way in which haloes are identified in simulations.
\item
The large-scale bias of $P^{\hc\hc}$ is not expected to be the same as
that of $P^{\delta\hc}$ due to nonlinear deterministic bias; this complicates studies of stochastic bias.
The difficulty of performing the halo-halo shot noise correction means
we are unable to make a strong statement about the non-equivalence of
$b^{\delta\hc}$ and $b^{\hc\hc}$.
\item 
As wavenumber increases, low mass haloes are increasingly anti-biased
and high mass haloes are increasingly positively biased.  Therefore,
the non-linear scale dependence of halo bias is not simply due to the
nonlinear evolution of the matter fluctuations.
\item 
Baryon acoustic oscillation features in the power spectrum are erased
on progressively larger scales as halo mass is increased.  In
addition, small shifts in the positions of the higher order peaks and
troughs occur which depend on halo mass.
\end{itemize}

In the second half of this paper we developed a `physical model' to
explain and reproduce these results. The model was constructed within
the framework of the halo model and we focused our attention on the
clustering of the halo centers. The halo-halo clustering term was
carefully propagated into the non-linear regime using `1-loop'
perturbation theory and a non-linear halo bias model. Our model can be
summarized as follows: The density field of haloes was assumed to be a
function of the local dark matter density field.  Under the condition
of small fluctuations, it was then expanded as a Taylor series in the
dark matter density with non-linear bias coefficients $b_i(M)$
\cite{FryGaztanaga1993}.  The density field was then evolved under 3rd
order Eulerian perturbation theory and this provided the 3rd order
Eulerian perturbed halo density field.  We then used the model to
derive the halo-halo and halo-dark matter cross power spectra up to
the 1-Loop level. This lead to the following conclusions:
\begin{itemize}
\item For $b_i$ non-vanishing, the effective bias on very, very large
scales for the halo spectra is not simply $b_1$, but also depends on
$b_2$, $b_3$ and the variance of fluctuations on scale $R$.  The halo
center power spectrum contains a term that corresponds to constant
power on very large scales. This implies that as $k\rightarrow0$, halo
bias should diverge as $[P_{\rm Lin}]^{-{1/2}}$.  The halo-dark matter
cross power spectrum does not exhibit this behavior. The predicted
bias from this statistic approaches a constant value on large scales.
\item When evaluated for a realistic cosmological model, with
non-linear bias parameters taken from the Sheth-Tormen mass function,
the theory is in broad agreement with numerical simulations for a wide
range of halo masses.
\item The non-linear evolution of the BAOs was also examined. The
model shows that non-linear bias and non-linear mode-mode coupling
increasingly damp the BAOs as halo mass is increased.  In addition,
the positions of the peaks and troughs can be shifted by small amounts
which depend on halo mass.
\end{itemize}
Using the ensemble of simulations we constructed scatter plots of the
halo versus dark matter over densities, contained in top-hat spheres
of size $R$ (see Appendix~\ref{app:MeasureBias}). From these, it was
shown that for filter scales $R<60h^{-1}\Mpc$ the bias was indeed
non-linear, and that, whilst the scatter increased as $R$ decreased,
the mean of the relationship did not change until $R<20h^{-1}\Mpc$. We
also examined whether the nonlinear bias parameters derived from the
Sheth \& Tormen mass function
\cite{ShethTormen1999,Scoccimarroetal2001} provided a reasonable match
to the empirical halo bias.

Using semi-empirical bias parameters as inputs for the analytic model
it was shown that the model well reproduced the scale dependence of
the halo-dark matter cross power spectra in the simulations, and for
all bins in halo mass. However, it was only qualitatively able to
reproduce the scale dependence of the non-linear halo bias.

The 1-Loop halo center power spectrum was then inserted into the halo
model framework and defined the `1-Loop Halo Model'. This was used to
predict the scale dependence of blue and red galaxy power
spectra. Plausible models for the Blue and Red Galaxy HODs were used
and the results showed complicated scale dependence.

Significant work still remains to be performed for this analytic
approach to be sharpened into a tool for precision cosmology. Some
possible improvements are: exchanging the 1-Loop matter power spectrum
for an accurate analytic fitting formula i.e. after the fashion of
{\tt halofit}, but designed purely for large scales; an application of
the new re-normalized perturbation theory techniques
\cite{CrocceScoccimarro2006a,CrocceScoccimarro2006b,McDonald2006a}
coupled with the non-linear bias model should certainly produce better
results.

The analytic and numerical results that we have developed here are
concerned purely with the clustering in real space. In a subsequent
paper we shall extend our analysis to explore the more realistic
situation of the scale dependence of dark matter, halo and galaxy
power spectra in redshift space.

\vskip 2pc

When this paper was nearing completion, a preprint appeared
\cite{McDonald2006b} with similar calculations of how nonlinear bias
changes the power spectrum.


\section*{Acknowledgements}

RES would like to thank Peter Schneider for a very thought
provoking discussion that has led to the current paper. We thank
Jacek Guzik, Gary Bernstein, Bhuvnesh Jain and Laura Marian for many
useful discussions during this work.  We would also like to thank
Pat McDonald, Daniel Eisenstein, Eiichiro Komatsu and Martin White for
useful comments on the draft. RES thanks Joerg Colberg and the Virgo
Consortium for providing access to the Hubble Volume and the VLS
simulations.  RES thanks CCPP and NYU for their kind hospitality
during part of this research. RS would like to thank M.~Manera for
useful discussions on halo exclusion. RKS would like to thank Martin
White for many illuminating discussions and the Aspen Center for
Physics. We thank M.~Crocce and S.~Pueblas for help regarding the
numerical simulations used here.  RES and RKS both acknowledge support
from the National Science Foundation under Grant No. 0520647. RS is
partially supported by NSF AST-0607747 and NASA NNG06GH21G.


\begin{widetext}

\appendix


\subsection{Poisson model} 

It is necessary to correct the measured dark matter power spectra for
shot-noise errors. These arise through approximating the continuous
CDM fluid by a point process.  If the discretization of the density
field obeys a Poisson process, i.e., dark matter particles are placed
with probability $\propto\rhob[1+\delta(\bx)]\delta V$, then the
`discreteness' correction is \cite{Peebles1980}:
\be P_{\rm true}(k)=P_{\rm obs}(k)-P_{\rm shot}\ ; \ \ P_{\rm
shot}=1/\nbar\ .\label{eq:Pdiscrete}\ee
Since the number of particles in our simulations is large, $512^3$,
this correction, on the scales of interest, is insignificant, e.g.
$P_{\rm shot}/P(0.1 h \Mpc^{-1})\sim10^{-3}$.  However, one must also
correct $P^{\hc\hc}$ for discreteness. If the dark matter haloes are
also regarded as a Poisson sampling of the smoothed halo density
field, then the correction will be the same but using the appropriate
number density $\nbar_{\rm h}$. In Figure \ref{fig:shotnoise} we show
the effect of this standard correction on our halo power spectra and
we plot the ensemble of the shot-noise corrected halo power. The
negative power values that result at high $k$ demonstrate that this
model must not be exactly correct and therefore for accurate
measurements it must be modified in some way. We now discuss a
possible explanation for this and propose a new shot correction.


\begin{figure*}
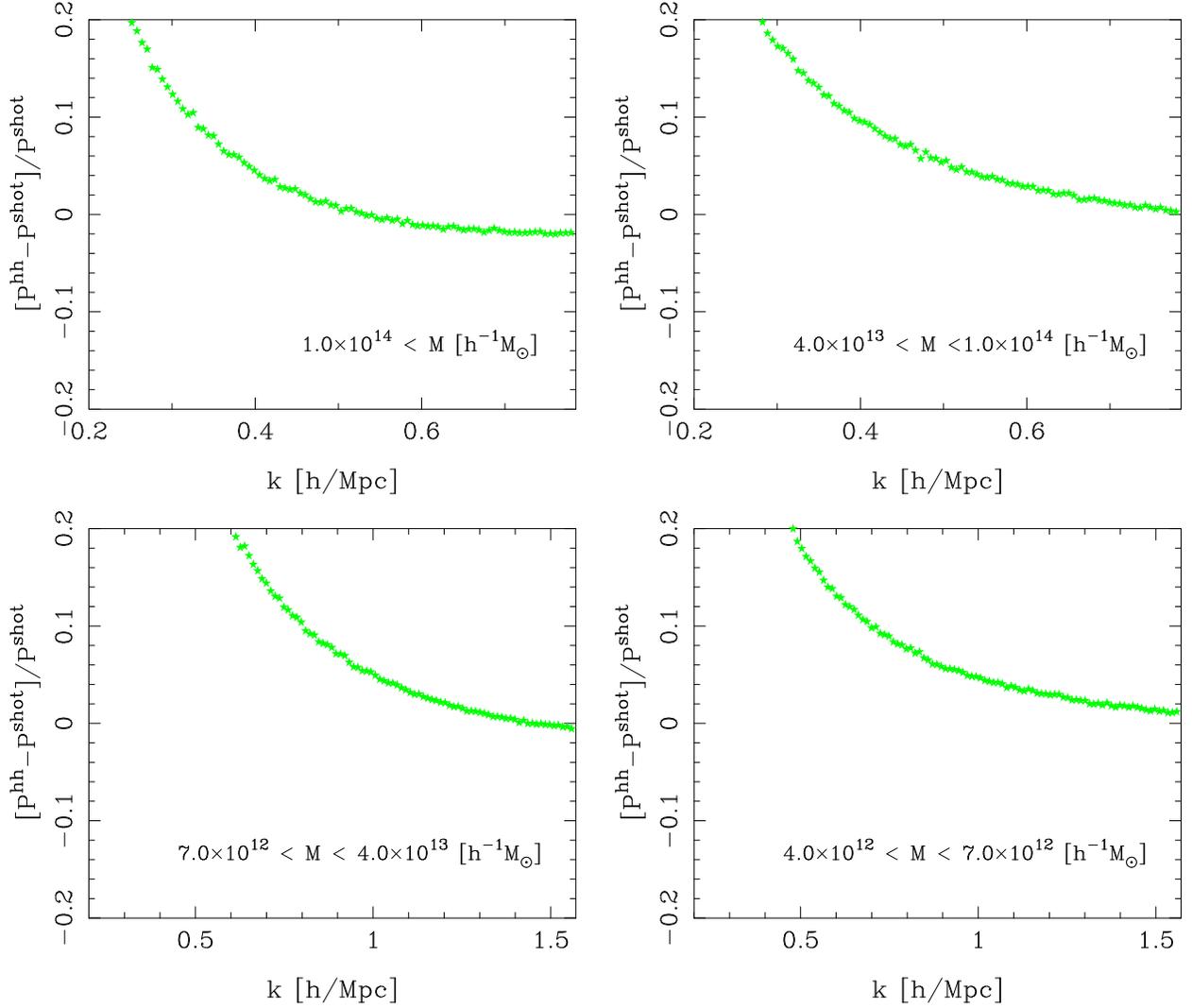

\centerline{
\includegraphics[width=8cm]{Fig.11a.ps}
\hspace{0.3cm}
\includegraphics[width=8cm]{Fig.11b.ps}}
\vspace{0.2cm}
\centerline{
\includegraphics[width=8cm]{Fig.11c.ps}
\hspace{0.3cm} 
\includegraphics[width=8cm]{Fig.11d.ps}}
\caption{\small{The effect of Poisson Shot noise correction on the
halo-halo power spectra for the four bins in halo mass. }
\label{fig:shotnoise}}
\end{figure*}


\section{Halo discreteness corrections}\label{app:HaloDiscrete}


\subsection{Halo exclusion effects}

As we have seen Poisson sampling must not be exact for haloes,
particularly those of large mass.  A possible reason is that by using
a friends of friends algorithm, one is automatically imposing that
haloes are never separated by distances smaller than about the sum of
their radii, else they would have been linked as a bigger halo.  As a
result of this exclusion effect, the two-point correlation function of
haloes drops dramatically from a value $\xi^h\gg 1$ at $r \sim r_e$ to
$\xi^h=-1$ for $r<r_e$, where $r_e$ is the exclusion radius. This
sharp drop of $\xi^h$ has an impact on the large-scale power spectrum,
as we now show.

Absent exclusion, the power spectrum would have been 
\be P^h(k) = \int_{0}^\infty \frac{\sin(kr)}{kr} \xi^h\, 4\pi r^2 dr.
\label{Pright}
\ee 
However, exclusion effects mean that 
\be
P^h(k) = \int_{r_e}^\infty \frac{\sin(kr)}{kr} \xi^h 4\pi r^2 dr 
            - v_{\rm TH}(k),
\label{Pwrong}
\ee
where $v_{\rm TH}(k)\equiv (4\pi/3) r_e^3 W_{\rm TH}(kr_e)$ with
$W_{\rm TH}$ the Fourier transform of a top-hat window in real space.
Hence, the difference in power due to exclusion is \be \Delta P^h(k) =
\int_{0}^{r_e} \frac{\sin(kr)}{kr}\,(1+\xi^h)\,4\pi r^2\, dr.  \ee
To estimate this, we must model $\xi^h$ at scales smaller than the
exclusion imposed by the friends of friends definition of a halo.  We
do so simply by approximating $\xi^h$ by a power-law $\xi^h = \xi_e
(r/r_e)^{-\gamma}$ obtained from fitting to the $r\ge r_e$
measurements, where $r_e$ is the exclusion scale.  In this
approximation
\be \Delta P^h(k) = v_{\rm TH}(k) + v(k)\ \xi_e,
\label{SPcorr}
\ee
where, in the large-scale approximation $kr_e\ll 1$,
\be v(k) \simeq \frac{4\pi r_e^3}{3-\gamma} - 
 \frac{2\pi r_e^3}{3(5-\gamma)} (kr_e)^2. \ee
Setting $\gamma=0$ gives $v_{\rm TH}$ in the same limit.  Thus, we see
that exclusion makes the halo power spectrum {\em smaller} by the
amount given by Eq.~(\ref{SPcorr}).  In contrast, Poisson shot noise
makes the power {\em larger} by $1/\nbar_{\rm h}$.  If Poisson noise
is subtracted from halo power spectra in simulations, then the result
becomes negative at high $k$.  So it may be that adding back the power
lost to exclusion will make the power positive again.  This is our
procedure. Note that the noise in Eq.~(\ref{SPcorr}) {\em is not white}. 

As a final note we reemphasize that all of these troubling issues may,
to a certain extent, be neatly side stepped, if we measure the
halo-dark matter cross power spectrum. The clear advantage of this
approach is that the `shot-noise' is dramatically reduced (there are
many more particles than haloes) and exclusion no longer plays a role.


\begin{figure}
\centerline{
\includegraphics[width=0.8\hsize,angle=0.0]{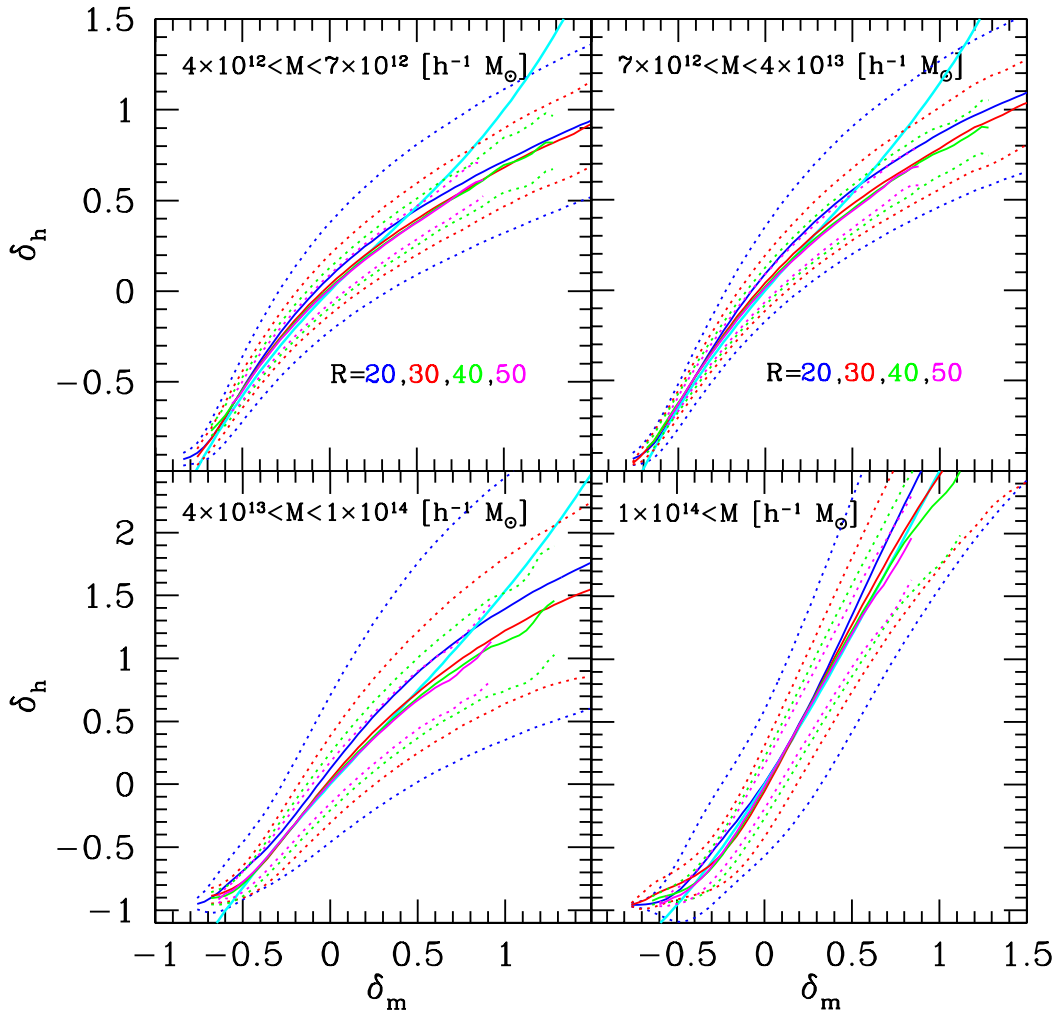}}
\caption{\small{Scatter plot of halo number-over-density vs the over
density of dark matter for four bins in halo mass. The fluctuations
were measured in top-hat spheres of radii $R_{\rm
TH}=\{20,30,40,50\}h^{-1}\Mpc$. The means and 1-sigma errors measured
from the ensemble are denoted by the solid blue, dash red, dot-dash
green and dotted magenta lines, respectively. The solid thick cyan
line shows the predictions from the Sheth \& Tormen model. The thick
triple dot-dash lines show the results from the cubic fit in $\delta$
as described in the text.}
\label{fig:scatterplot}}
\end{figure}


\section{Measuring the non-linear bias parameters}
\label{app:MeasureBias}

The accuracy of 1-Loop Halo PT depends on the accuracy of the halo
bias parameters $b_1$, $b_2$ and $b_3$.
Figure~\ref{fig:halobiasparams} shows these parameters derived from
the conditional Sheth \& Tormen mass function
\cite{Scoccimarroetal2001}.  There are a number of ways to test the
accuracy of the bias parameters.  The most direct is to smooth the
halo and mass fields with a filter of scale $R$, and to then plot
$\delta^{\hc}(x|R)$ versus $\delta(x|R)$ in the smoothed fields
(e.g. \cite{ShethLemson1999,CasasMirandaetal2002}).  From the
resulting scatter plot, one then directly fits for the parameters
$b_i$ up to some order, subject to the constraint that
$\left<\delta^{\hc}\right>=\left<\delta\right>=0$.

Figure \ref{fig:scatterplot} shows such scatter plots for the four
bins in halo mass described earlier (Table \ref{table:halocat}).  In
all cases, unlike elsewhere in the paper, the fields were filtered
with a `top-hat' window function.  Solid blue, red, green and magenta
lines show results for four filter scales $R_{\rm TH}=\{20,30,40,50\}
h^{-1}\Mpc$. The upper and lower dotted lines show the $1-\sigma$
errors about the mean relation (solid lines).  (Smoothing with a Gaussian filter instead does not
significantly change the mean, but does slightly change the scatter
around the mean relation.)  The figure shows that, above a certain
scale, $\left<\delta^{\hc}|\delta\right>$ defines a universal curve
which is independent of $R$; it is this curve which the bias
coefficients are supposed to describe.  The figure also shows that the
scatter around this mean curve decreases as $R$ increases.  The
assumption that bias is deterministic is equivalent to assuming that
this scatter, this stochasticity, is negligible.  Clearly, this is a
reasonable approximation only for large smoothing scales $R$, and $R$
must be bigger for the rarer, more massive haloes before the bias can
be called deterministic.

The main problem with this approach is that, when $R$ is large, then
the distribution of $\delta$ becomes sharply peaked about its mean
value (zero!), so estimating the higher order bias coefficients
becomes difficult---only the linear bias term can be reliably
measured.  In effect, determination of the bias coefficients
corresponds to fitting a polynomial to these mean curves, and the
best-fitting coefficients will be correlated.  The results of this
exercise are tabulated in \ref{table:halobias}, which also shows the
values of $b_i$ predicted from the Sheth-Tormen mass function.  Some
of the discrepancy is a consequence of the fact that the Sheth-Tormen
mass function under-predicts the abundance of high mass haloes in the
HR and LR simulations by up to 20\% (see
\cite{CroccePeublasScoccimarro2006} for an explanation as to why this
happens).  But note that, because of the correlation between the
fitted coefficients, the discrepancy with the theory values is
difficult to assess.  The solid thick (cyan) line in the figure shows
the mean bias relation derived from the Sheth-Tormen mass function.

In light of these issues we use the following prescription when
generating the theory curves: $b_1$ coefficients are to be measured
directly from the data as described in Section
\ref{ssec:halopowerspectra} and the parameters $b_2$ and $b_3$ are to
be derived through computing the following integrals over the relevant
bin width:
\be \overline{b}_i(M)=\frac{1}{\nbar(M_1,M_2)}
\int_{M_1}^{M_2}dM n(M) b_i(M) \ ,\ee
where $n(M)$ and $b_i(M)$ are the Sheth \& Tormen mass function and
bias parameters. We shall refer to this approach as the semi-empirical
method. It shall be left to our future work to to provide a more
self-consistent solution to this problem.


\section{Eulerian PT}\label{app:PTkernels}

\subsection{Real space kernels}

The growth of density inhomogeneities in an expanding universe may be
explored in the single stream approximation using the Eulerian fluid
equations. In Fourier space the equations governing the evolution of
the fluctuations to the density ($\delta(\bk,t)$) and divergence of
the peculiar velocity field ($\theta(\bk,t)\equiv\nabla\cdot{\bf v}$)
are written \citep{Bernardeauetal2002}:
\ba & & \frac{\partial\delta(\bk,t)}{\partial t}+\theta(\bk,t) = -\int
\frac{\dq_1 \dq_2}{(2\pi)^3} \left[\delta^D(\bk)\right]_{2}
\alpha(\bq_1,\bq_2) \theta(\bq_1,t)\delta(\bq_2,t) \\ & &
\frac{\partial\theta(\bk,t)}{\partial t}+H(a)\theta(\bk,t)+\frac{3}{2}
\Omega_m H^2(t)\delta(\bk,t) = -\int \frac{\dq_1 \dq_2}{(2\pi)^3}
\left[\delta^D(\bk)\right]_{2}
\beta(\bq_1,\bq_2)\theta(\bq_1,t)\theta(\bq_2,t) \ea
where 
\be \alpha_{i,j}\equiv \alpha(\bq_{i},\bq_{j})
\equiv \frac{\bk\cdot\bq_i}{q_i^2}\ ;
\hspace{0.3cm} \beta_{i,j}\equiv\beta(\bq_i,\bq_j)\equiv 
\frac{k^2(\bq_i\cdot\bq_j)}{2q_i^2q_j^2}\ ;\hspace{0.3cm} \bk=\bq_i+\bq_j\ .
\ee
As shown in \cite{Bernardeauetal2002}, for the case of the Einstein-de
Sitter model the density and velocity divergence fields may be
expanded as a perturbation series and the solutions at each order may
then be written down in terms of the density and velocity
perturbations from all lower orders: e.g.
\ba 
& & 
\delta(\bk)=\sum_{n=1}^{\infty} D_1^{n}(a) \delta_n(\bk)
\ ;\hspace{2cm}
\delta_n(\bk)=\int
\frac{\prod_{i=1}^{n}\left\{\dk_i\,\delta_1(\bk_i)\right\}}{(2\pi)^{3n-3}}
\left[\delta^D(\bk)\right]_n F^{(s)}_n(\bk_1,...,\bk_n) \label{eq:PTdelta}
\\
& & 
\theta(\bk)=H(a)f(a)\sum_{n=1}^{\infty} D_1^{n}(a) \theta_n(\bk) 
\ ;\hspace{0.5cm}
\theta_n(\bk)=\int
\frac{\prod_{i=1}^{n}\left\{\dk_i\,\delta_1(\bk_i)\right\}}{(2\pi)^{3n-3}}
\left[\delta^D(\bk)\right]_n G^{(s)}_n(\bk_1,...,\bk_n)
\label{eq:PTdivvel}\ ,\ea
where $H(a)$ is the dimensionless Hubble parameter, $D_1(a)=a$ is the
linear growth factor and $f(a)=d \log D/d \log a
\approx\Omega_m^{5/9}$ gives the velocity growth suppression
factor. The functions $F^{(s)}_n$ and $G^{(s)}_n$ are the PT kernels
for $\delta$ and $\theta$ symmetrized in all of their arguments,
respectively. The notation
$\left[\delta^D(\bk)\right]_n=\delta^D(\bk-\bq_1-\dots-\bq_n)$ was
adopted.
The first three symmetrized density kernels are:
\ba 
& & \Fsym_1(\bq)=1 \ ; \\
& & \Fsym_2(\bq_1,\bq_2)=\frac{5}{14}\left[\alpha_{1,2}+\alpha_{2,1}\right]
+\frac{2}{7}\beta_{1,2} \ ;\\
& & \Fsym_3(\bq_1,\bq_2,\bq_3)=\frac{7}{54}
\left[\Fsym_{1,2}\alpha_{3,12}+\Fsym_{2,3}\alpha_{1,23}+\Fsym_{3,1}\alpha_{2,31}
+\Gsym_{1,2}\alpha_{12,3}+\Gsym_{2,3}\alpha_{23,1}+\Gsym_{3,1}\alpha_{31,2}
\right]\nonumber\\
& & \hspace{2.5cm}+\frac{2}{27}\left[
\Gsym_{1,2}\beta_{12,3}+\Gsym_{2,3}\beta_{23,1}+\Gsym_{3,1}\beta_{31,2}
\right]
\ .\ea
The first three symmetrized velocity divergence PT kernels are:
\ba 
& & \Gsym_1(\bq)=1 \ ; \\
& & \Gsym_2(\bq_1,\bq_2)=\frac{3}{14}\left[\alpha_{1,2}+\alpha_{2,1}\right]
+\frac{4}{7}\beta_{1,2}\ ; \\
& & \Gsym_3(\bq_1,\bq_2,\bq_3)=\frac{1}{18}
\left[\Fsym_{1,2}\alpha_{3,12}+\Fsym_{2,3}\alpha_{1,23}+\Fsym_{3,1}\alpha_{2,31}
+\Gsym_{1,2}\alpha_{12,3}+\Gsym_{2,3}\alpha_{23,1}+\Gsym_{3,1}\alpha_{31,2}
\right] \nonumber \\
& & \hspace{2.7cm}
+\frac{2}{9}\left[
\Gsym_{1,2}\beta_{12,3}+\Gsym_{2,3}\beta_{23,1}+\Gsym_{3,1}\beta_{31,2}
\right]
\ .\ea
where we have adopted the compact notation
\be 
\Fsym_{i_1,\dots,i_n}\equiv \Fsym_n(\bq_{i_1},\dots,\bq_{i_n})\ ;\hspace{0.5cm}
\Gsym_{i_1,\dots,i_n}\equiv \Gsym_n(\bq_{i_1},\dots,\bq_{i_n})
\ee
and
\be 
\alpha_{i_1 \dots i_n,j_1\dots j_m}\equiv
\alpha(\bq_{i_1}+\dots+\bq_{i_n},\bq_{j_1}+\dots+\bq_{j_m})\ ; \hspace{0.5cm}
\beta{i_1 \dots i_n,j}\equiv
\beta(\bq_{i_1}+\dots+\bq_{i_n},\bq_{j})\ .
\ee 

Exact analytic solutions for arbitrary cosmological models have not
yet been found. However, as was shown in \cite{Scoccimarroetal1998},
under the assumption that $D_n(a)\propto [D_1(a)]^n$ and
$f(\Omega)\approx\Omega_m^{0.5}$, the solutions are identical to the
Einstein-de Sitter solutions, but the growth factors are changed to
those for the particular cosmological model in question. All other
changes are small corrections \cite{Bernardeauetal2002}.


\subsection{Evaluation of 1-Loop expressions}\label{app:1loop}

Some care is required in the numerical evaluation of the Halo-PT
expressions because fine cancellations can occur between negative and
positive terms. The approach that we adopt throughout this paper can
be demonstrated through the following example.  Consider the 1-Loop
power spectrum for CDM given by equation (\ref{eq:1Loop}). The PT
corrections $P_{13}$ and $P_{22}$ maybe analytically developed up to
the following points:
\ba 
& & \hspace{-1cm}P_{13}(k) = 
6P_{11}(k)\int\frac{dq}{(2\pi)^3}q^2P_{11}(\bq)\int d\hat{\bq}\ 
F_3(\bk,\bq,-\bq) \nonumber 
\\
& & \hspace{0.2cm} = \frac{P_{11}(k)k^3}{252(2\pi)^2}\int_{0}^{\infty} dx\, x^2 
P_{11}(xk)\left\{-42x^{2}+100-\frac{158}{x^{2}}+\frac{12}{x^{4}}+
\frac{3}{x}(1-x^2)^3(7x^2+2)\log\left[\frac{x+1}{|x-1|}\right]\right\} ;
\\
& & \hspace{-1cm} P_{22}(k) = 2\int \frac{\dq}{(2\pi)^3}P_{11}(q)P_{11}(|\bk-\bq|) 
\left[F_{2}(\bq,\bk-\bq)\right]^2 \nonumber 
\\
& & \hspace{0.2cm} = 2 \int_0^{\infty} \frac{dq}{(2\pi)^2} dq P_{11}(q)
\int_{-1}^{1} d\mu P_{11}(k\psi(x,\mu))
\left\{\frac{5}{7}+\frac{1}{2}\frac{\mu-x}{\psi(x,\mu)}
\left[\frac{x}{\psi(x,\mu)}+\frac{\psi(x,\mu)}{x}\right]
+\frac{2}{7}\left[\frac{\mu-x}{\psi(x,\mu)}\right]^2\right\}^2\ ,
\ea
where $x=q/k$ and where $\psi^2(x,\mu)=1+x^2-2x\mu$. If the $P_{13}$
integral is truncated on large and small scales to inhibit infra-red
and ultra-violet divergences, as may occur for some power spectra,
then identical constraints must also be placed on the $P_{22}(k)$.
Explicitly, if we adopt
\be P(q)=0 \hspace{0.5cm} {\rm for}\ \  \left\{ \begin{array}{l}
k<k_{\rm fun}\nonumber\\
k>k_{\rm cut}
\end{array}\right. \ ,
\ee
then the angular integral for $P_{22}$ must necessarily have the new
limits
\be 
\int_{-1}^{1}d\mu\rightarrow\int_{\mu_1}^{\mu_2}\ ;
\hspace{0.5cm} 
\mu_2=\min\left[1,\sqrt{(k^2+q^2-k^2_{\rm cut})/2kq}\right] \ ; \hspace{0.5cm}
\mu_1=\max\left[-1,\sqrt{(k^2+q^2-k^2_{\rm fun})/2kq}\right]\ .
\ee
Lastly, the particular values for the limits $k_{\rm fun}$ and $k_{\rm
cut}$ were selected so that variance integrals, $\sigma^2(R)$, would
be convergent within the finite range.

\end{widetext}



\end{document}